\documentclass[letterpaper,journal]{IEEEtran}
\usepackage{amsmath,amsfonts,amssymb,mathtools,amsthm}
\usepackage{array,tabularray}
\UseTblrLibrary{booktabs}
\usepackage[caption=false,font=normalsize,labelfont=sf,textfont=sf]{subfig}
\usepackage{graphicx}
\usepackage{cite}
\usepackage{bm}
\usepackage{siunitx}
\usepackage{mathrsfs}
\usepackage{url}
\usepackage{placeins}
\usepackage{hyperref}
\usepackage[capitalise,nameinlink]{cleveref}
\usepackage{orcidlink}

\newtheorem{proposition}{Proposition}[section]

\crefname{section}{Sec.}{Secs.}
\crefname{subsection}{Sec.}{Secs.}
\crefname{subsubsection}{Sec.}{Secs.}
\crefname{figure}{Fig.}{Figs.}
\crefname{table}{Tab.}{Tabs.}
\crefname{equation}{Eq.}{Eqs.}
\hypersetup{hidelinks}

\begin{document}

\title{First-Order Observation Modeling of Reference--Background Residuals in Single-Snapshot FDA-MIMO-GPR}

\author{Yisu Yan$^{\orcidlink{0009-0008-4022-3619}}$,~\IEEEmembership{Graduate Student Member,~IEEE,} Jifeng Guo$^{\orcidlink{0000-0002-9710-0045}}$ % ,~\IEEEmembership{Member,~IEEE}

	\thanks{Yisu Yan is with School of Astronautics, Harbin Institute of Technology, Heilongjiang, China
		(e-mail: \href{mailto:23B918085@stu.hit.edu.cn}{23B918085@stu.hit.edu.cn}).}

	\thanks{Jifeng Guo is with School of Astronautics, Harbin Institute of Technology, Heilongjiang, China (e-mail: \href{mailto:guojifeng@hit.edu.cn}{guojifeng@hit.edu.cn}).}

	\thanks{This work has been submitted to the IEEE for possible publication. Copyright may be transferred without notice, after which this version may no longer be accessible.}
}

\maketitle

\begin{abstract}
	This paper develops a first-order observation model for reference--background mismatch in single-snapshot frequency-diverse array multiple-input multiple-output ground-penetrating radar (FDA-MIMO-GPR) over complex media. The model separates the shared background residual from the local anomaly and maps it through a fixed-reference distorted Born approximation into the joint observation domain. The formulation distinguishes the physical residual source from its organization by FDA frequency states and transmit--frequency assignments. Closed-form residual moments are transferred to fixed linear task domains, yielding task-risk and discrepancy relations together with a finite reference-screening criterion. Controlled Cole--Cole ensembles reveal nonzero off-event covariance and task-dependent pseudo-anomaly patterns. Finite full-wave simulations support the predicted covariance and mismatch-dependent discrepancy trends while delimiting the first-order validity domain. These results provide a compact framework for quantifying and screening reference-induced residuals in single-snapshot FDA-MIMO-GPR.
\end{abstract}

\begin{IEEEkeywords}
	ground-penetrating radar, FDA-MIMO, single-snapshot observation, distorted Born approximation, reference medium, reference--background residual, channel-block covariance
\end{IEEEkeywords}

\section{Introduction}\label{sec:intro}

\IEEEPARstart{G}{round-penetrating} radar (GPR) observations contain both target responses and contributions from the host medium. This study considers a single-snapshot frequency-diverse array multiple-input multiple-output ground-penetrating radar (FDA-MIMO-GPR) acquisition, in which multiple transmit--receive channels are acquired under prescribed transmitter-dependent frequency offsets. In this setting, permittivity, conductivity, water content, dispersion, attenuation, and interface structure affect the phase and amplitude of the received field \cite{topp1980electromagneticSoilWater,vandam2014calibrationFunctionsGPR,zajicova2019gprSoilReview}. Frequency-dependent constitutive models, including Cole--Cole descriptions and their relation to Debye decompositions, provide a parameterization of these medium effects \cite{tarasov2013coleColeSIP,weigand2016debyeDecompositionSIP,weigand2016relationshipColeCole}.

Complex-medium GPR formulations commonly describe incremental scattering relative to a prescribed background state. The distorted Born approximation (DBA) provides a first-order propagation model for this purpose. Its incident field and Green tensor are evaluated at the prescribed state, and the incremental contrast is mapped to the receiver through the resulting propagation kernel \cite{soldovieri2004dbaSlab,meincke2001linearGprInversion}.

Existing treatments differ in the quantity that they hold fixed, update, or transfer. Fixed-reference DBA and linearized GPR formulations hold the background-dependent kernel fixed. Distorted Born iterative and modified-gradient methods update the kernel or gradient during reconstruction \cite{chew1990dbimPermittivity,kleinman1992modifiedGradient,belkebir2001modifiedGradientBorn,cui2001dbimLossyEarth}. Nonlinear Born extensions retain higher-order field relations \cite{habashy1993beyondBornRytov}. Contrast source inversion (CSI) and GPR full-waveform inversion (GPR-FWI) update contrast, medium, or source descriptions through iterative objectives \cite{berg1997contrastSource,berg2001csiStateOfArt,abubakar2008fdCsi,guo2020idMrcsi,lambot2004modelingGpr,qin2022freqDependentGprFwi,qin2022jointPetrophysicalGpr,tan2025multiobjectiveGprFwi}. Image and path compensation methods address propagation effects in a subsequent operator domain \cite{zhou2025scfbp,wang2026compensationRtm}.

\begin{table*}[htbp]
	\centering
	\caption{Comparison of background-mismatch treatments relevant to the present scope.}
	\label{tab:intro_literature_comparison}
	\begin{tblr}{
			width=\textwidth,
			colspec={X[1.1] X[1.2] X[1.2] X[0.8] X[1.2]},
			row{1}={font=\bfseries},
			cells={font=\footnotesize}
		}
		\toprule
		Method family                                                                                             & Reference treatment                                                                  & Mismatch handling                                               & Primary output                  & Relation to this paper                                                                    \\
		\midrule
		Fixed-reference DBA and linearized inversion \cite{soldovieri2004dbaSlab,meincke2001linearGprInversion}   & A background or reference kernel is held fixed for a first-order scattered-field map & Mismatch remains in modeled-data or reconstruction error        & Target or contrast estimate     & Provides the fixed-kernel propagation starting point used here                            \\
		Iterative Born and modified-gradient methods \cite{chew1990dbimPermittivity,kleinman1992modifiedGradient} & Kernel or background-dependent gradient is updated                                   & Mismatch is reduced through iterative field or contrast updates & Updated reconstruction          & Uses an update mechanism distinct from the fixed statistical analysis here                \\
		Nonlinear Born extensions \cite{habashy1993beyondBornRytov}                                               & Field relation is extended beyond Born or Rytov linearization                        & Higher-order scattering enters a nonlinear formulation          & Nonlinear inverse estimate      & Defines a different validity branch                                                       \\
		Contrast-source, CSI, and GPR-FWI \cite{berg1997contrastSource,qin2022freqDependentGprFwi}                & Contrast sources, medium parameters, or waveform models are updated                  & Mismatch is absorbed by an iterative objective                  & Medium or source reconstruction & Transfers information through an inversion objective; this study keeps the task map fixed \\
		GPR path and image correction \cite{zhou2025scfbp,wang2026compensationRtm}                                & Propagation or imaging operators are corrected                                       & Phase, attenuation, or path errors are compensated              & Corrected waveform or image     & Operates in a different task domain                                                       \\
		\bottomrule
	\end{tblr}
\end{table*}

\Cref{tab:intro_literature_comparison} groups prior work according to the quantity that is held fixed, updated, or transferred. These treatments therefore place background mismatch in different objects and domains. The present study retains the fixed propagation kernel and examines the residual left by its reference choice as a statistical object in the joint observation and fixed-task domains.

FDA-MIMO studies characterize transmitter-dependent frequency organization through range--angle coupling, single-snapshot estimation, coherent integration, and frequency-increment design \cite{xiong2018fdaMimoRangeAngle,lan2021singleSnapshotFdaMimo,miao2025frequencyCoupling,lan2025frequencyIncrement,qi2025waveformDesignFdaMimo}. A subsurface FDA-MIMO radar study provides a related propagation setting with a target-detection formulation \cite{liu2018detectionSubsurfaceFDAMIMO}. This observation-side literature supplies the frequency-state and transmit--frequency organization needed for the scope comparison in \cref{tab:intro_scope_comparison}.

\begin{table*}[htbp]
	\centering
	\caption{Scope comparison between existing treatments and the present FDA-MIMO-GPR formulation.}
	\label{tab:intro_scope_comparison}
	\begin{tblr}{
			width=\textwidth,
			colspec={X[1.0] X[1.1] X[1.1] X[1.2] X[1.2]},
			row{1}={font=\bfseries},
			cells={font=\footnotesize}
		}
		\toprule
		Literature line                                                                                          & Observation organization                              & Medium treatment                                          & Mismatch object and output                                            & Connection to this paper                                      \\
		\midrule
		Fixed-reference DBA \cite{soldovieri2004dbaSlab,meincke2001linearGprInversion}                           & Background-dependent scattered-field map              & Fixed background or reference kernel                      & Target or contrast reconstruction                                     & The present residual is defined within this branch            \\
		DBIM, CSI, and GPR-FWI \cite{chew1990dbimPermittivity,berg1997contrastSource,qin2022freqDependentGprFwi} & Iterative multichannel or multioffset waveform data   & Kernel, contrast source, or medium parameters are updated & Updated contrast or medium estimate                                   & Different update objective                                    \\
		GPR path and loss compensation \cite{zhou2025scfbp,wang2026compensationRtm}                              & GPR-SAR or waveform-to-image operator                 & Propagation path, relaxation, or loss model               & Phase, attenuation, or image correction                               & Different correction domain                                   \\
		FDA-MIMO and subsurface FDA radar \cite{xiong2018fdaMimoRangeAngle,liu2018detectionSubsurfaceFDAMIMO}    & Transmitter-dependent offsets                         & Radar steering and target model                           & Range, angle, detection, or integration output                        & Cole--Cole GPR residual is outside the cited radar model      \\
		Current work                                                                                      & Single-snapshot FDA event vector with GPR propagation & Fixed reference and first-order DBA branch                & Shared residual, covariance, task transfer, and conditional screening & Scope-specific combination with an explicit validity boundary \\
		\bottomrule
	\end{tblr}
\end{table*}

\Cref{tab:intro_scope_comparison} places the present formulation at the intersection of fixed-reference GPR propagation and transmitter-dependent FDA-MIMO observation organization. The reference medium is the selected working state used to construct the propagation kernel. The physical background is the actual host medium after removal of the local anomaly. The true medium contains the physical background and the local anomaly. Their parameter difference is termed the reference--background parameter residual. Its fixed-reference first-order response is the shared residual response. Kernel variation, higher-order constitutive effects, multiple scattering, interface feedback, target--background coupling, and instrument or modeling disturbances are collected in the modeling remainder.

The combination gap concerns how the reference--background parameter residual propagates through a single-snapshot FDA-MIMO-GPR observation and a downstream task operator. The residual is the physical source of common event dependence. FDA frequency states and transmit--frequency assignments act through $\mathbf L(\mathbf Z)$ to organize its eventwise propagation and covariance. A zero frequency increment still permits common residual structure in a conventional multichannel observation. The study therefore traces the residual from complex-medium propagation, through the joint observation domain, to a fixed task domain, where its consequences are expressed as task risk an, for spatial task maps, a pseudo-anomaly component.

The paper makes the following contributions within this scope:
\begin{enumerate}
	\item A fixed-reference, first-order single-snapshot FDA-MIMO-GPR propagation and observation model is established for complex Cole--Cole media.
	\item The reference--background parameter residual is distinguished from the local anomaly, exact full-wave background difference, modeling remainder, and noise.
	\item The joint observation mean and covariance are derived, including off-event blocks and their dependence on FDA frequency states and transmit--frequency assignments.
	\item The residual moments are transferred to fixed linear task domains, where the resulting moments are expressed as task risk and a task-dependent spatial pseudo-anomaly component.
	\item A finite reference-state screening interface is formulated and evaluated under the admitted first-order domain, without interpreting screening as material identification or parameter recovery.
\end{enumerate}

The remainder of the paper is organized as follows. \cref{sec:signal_model} develops the complex-medium fixed-reference propagation and observation model. \cref{sec:observation_statistics} derives the residual response and its FDA-dependent joint covariance. \cref{sec:task_transfer} transfers these quantities to fixed linear task domains and defines the finite reference-screening interface. \cref{sec:experiments} reports the controlled analytical experiments and finite full-wave checks. \cref{sec:conclusion} states the resulting scope and future directions.

\section{Fixed-Reference FDA-MIMO-GPR Propagation and Observation Model}\label{sec:signal_model}

The principal symbols used throughout the model and experiments are summarized in \cref{tab:notation}. Acronyms are defined at first use and are not repeated in this table.

\begin{table*}[!t]
	\centering
	\caption{Notation used in the propagation, covariance, and task-domain models.}
	\label{tab:notation}
	\begin{tblr}{
			width=\textwidth,
			colspec={X[1.1] X[2.9]},
			row{1}={font=\bfseries},
			cells={font=\scriptsize}
		}
		\toprule
		Symbol                                                                               & Definition                                                                                                                                 \\
		\midrule
		$\bm\mu$                                                                             & $[\varepsilon_\infty,\Delta\varepsilon,\tau,\alpha,\sigma]^{\mathrm T}$                                                                    \\
		$\bm\mu_b$, $\bm\mu_{\mathrm{ref}}$, $\bm\mu_{\mathrm{true}}$                        & physical background, reference medium, and complete medium                                                                                 \\
		$\delta\bm\mu$, $\delta\bm\mu_{\mathrm{an}}$, $\delta\bm\mu_{\mathrm{sh}}$           & total, anomaly, and reference--background parameter differences                                                                            \\
		$\varepsilon_c$, $\mathcal M_{\mathrm{CC}}$                                          & complex permittivity and admissible Cole--Cole state set                                                                                   \\
		$\mathbf B_e$, $\mathbf H_e$, $\mathbf L_e$, $\mathbf L$                             & constitutive, propagation, event, and joint tangent operators with $\mathbf L_e=\mathbf H_e\mathbf B_e$                                    \\
		$\mathbf Z$, $K_f$, $\omega_e$                                                       & one-hot event assignment, number of frequency states, and assigned event frequency                                                         \\
		$\bm c^{(\mathrm{sh})}$, $\bm c^{(\mathrm{sh,ex})}$, $\bm r^{(\mathrm{sh})}$         & first-order shared response, exact background-state difference, and its remainder                                                          \\
		$\bm y$, $\bm n$                                                                     & joint observation and instrumental/noise term                                                                                              \\
		$\mathbf C_\mu$, $\mathbf R_{\mathrm{sh}}$                                           & parameter covariance and shared-residual observation covariance with $\mathbf R_{\mathrm{sh}}=\mathbf L\mathbf C_\mu\mathbf L^{\mathrm H}$ \\
		$\mathbf R_{\mathrm{full}}$, $\mathbf R_{\mathrm{blk}}$, $\mathbf R_{\mathrm{diag}}$ & full, event-block, and diagonal covariance models                                                                                          \\
		$\mathbf A$, $\bm e_A^{(\mathrm{sh})}$, $J_A^{(\mathrm{sh})}$                        & fixed task operator, shared-residual task component, and task risk                                                                         \\
		$N_e$, $N_r$, $P$, $P_A$, $q$                                                        & event count, receiver count, voxel count, task dimension, and Cole--Cole component index; $q=1,\ldots,5$                                   \\
		$D_{\mathrm{obs}}$, $D_A$, $\rho$, $\Delta f$, $h$                                   & observation/task discrepancy, mismatch coordinate, frequency increment, and tangent step                                                   \\
		\bottomrule
	\end{tblr}
\end{table*}

\subsection{Complex-medium states and reference--background decomposition}\label{subsec:constitutive}

Let the local complex-medium state at position $\bm{x}$ be represented by the five-parameter Cole--Cole vector
\begin{equation}
	\bm{\mu}(\bm{x}) = \left[\varepsilon_{\infty}(\bm{x}),\Delta\varepsilon(\bm{x}),\tau(\bm{x}),\alpha(\bm{x}),\sigma(\bm{x})\right]^{\mathrm T}.
	\label{eq:mu_def}
\end{equation}
The corresponding complex permittivity is written as
\begin{equation}
	\varepsilon_c(\omega,\bm{x}) = \varepsilon_\infty(\bm{x}) + \frac{\Delta\varepsilon(\bm{x})}{1+(\mathrm{i}\omega\tau(\bm{x}))^{1-\alpha(\bm{x})}} - \frac{\mathrm{i}\sigma(\bm{x})}{\omega\varepsilon_0}.
	\label{eq:eps_local}
\end{equation}
The admissible Cole--Cole set is
\begin{equation}
	\mathcal M_{\mathrm{CC}} =
	\left\{\bm{\mu}\,\middle|\,
	\begin{aligned}
		 & \varepsilon_\infty>0,\quad \Delta\varepsilon\geq 0,\quad \tau>0, \\[-1mm]
		 & 0\leq\alpha<1,\quad \sigma\geq0
	\end{aligned}\right\}.
	\label{eq:cole_cole_admissible_set}
\end{equation}
All states used below are required to produce finite passive responses in the analysis band. The reference state is also required to have a nonzero complex permittivity in that band, expressed by
\begin{equation}
	\inf_{\substack{\bm{x}\in\Omega\\\omega\in\Omega_{\mathrm{op}}}}
	\left|\varepsilon_{\mathrm{ref}}(\omega,\bm{x})\right|>0.
	\label{eq:reference_permittivity_nonzero}
\end{equation}

Let $\bm{\mu}_{\mathrm{true}}$ denote the complete medium, $\bm{\mu}_b$ the physical background with the local anomaly removed, and $\bm{\mu}_{\mathrm{ref}}$ the selected reference medium. The local anomaly increment is
\begin{equation}
	\delta\bm{\mu}_{\mathrm{an}} = \bm{\mu}_{\mathrm{true}}-\bm{\mu}_b,
	\label{eq:delta_mu_an}
\end{equation}
Equivalently,
\begin{equation}
	\bm{\mu}_{\mathrm{true}}=\bm{\mu}_b+\delta\bm{\mu}_{\mathrm{an}}.
	\label{eq:true_bg_an_relation}
\end{equation}
and the total parameter difference from the reference is
\begin{equation}
	\delta\bm{\mu}=\bm{\mu}_{\mathrm{true}}-\bm{\mu}_{\mathrm{ref}},\qquad
	\delta\bm{\mu}_{\mathrm{sh}}=\bm{\mu}_b-\bm{\mu}_{\mathrm{ref}}.
	\label{eq:delta_mu}
\end{equation}
The shared residual field is therefore
\begin{equation}
	\delta\bm{\mu}_{\mathrm{sh}}(\bm{x})
	=\bm{\mu}_b(\bm{x})-\bm{\mu}_{\mathrm{ref}}(\bm{x}).
	\label{eq:delta_mu_sh_field}
\end{equation}
Consequently,
\begin{equation}
	\delta\bm{\mu}=\delta\bm{\mu}_{\mathrm{sh}}+\delta\bm{\mu}_{\mathrm{an}}.
	\label{eq:delta_mu_split}
\end{equation}
Equation~\eqref{eq:delta_mu_split} is an identity in parameter space. Complete electromagnetic fields remain governed by nonlinear constitutive and propagation maps. Field-level additivity is introduced only for the fixed-reference first-order model below.

\subsection{Reference propagation kernel and first-order event response}\label{subsec:reference_kernel}\label{subsec:channel_model}

An event is a transmit-coded channel separated by frozen receiver processing within one coherent acquisition. Events are not independent temporal snapshots. For event $e$, let $\omega_e$ denote its assigned frequency and let $\mathcal L_{\mathrm{ref}}(\omega_e)$ be the Maxwell operator at the reference state. Its Green tensor and reference incident field are denoted by $\mathbf G_{\mathrm{ref}}(\omega_e;\bm{x},\bm{x}')$ and $\mathbf E_{\mathrm{inc},e}^{\mathrm{ref}}(\bm{x})$. Receiver processing, geometry, polarization, and sampling are included in the effective reference measurement kernel $\mathbf K_e^{(0)}$.

Under the time convention $e^{j\omega t}$, the frequency-domain Maxwell operator associated with the reference medium is
\begin{equation}
	\mathcal L_{\mathrm{ref}}(\omega)\bm E
	\coloneqq
	\nabla\times\nabla\times\bm E
	-\omega^2\mu_0\varepsilon_{\mathrm{ref}}(\omega,\bm{x})\bm E,
	\label{eq:ref_operator}
\end{equation}
where
\begin{equation}
	\varepsilon_{\mathrm{ref}}(\omega,\bm{x})
	\coloneqq
	\mathcal F_{\mathrm{CC}}\bigl(\omega;\bm{\mu}_{\mathrm{ref}}(\bm{x})\bigr).
	\label{eq:eps_ref}
\end{equation}
The corresponding Green tensor satisfies
\begin{equation}
	\mathcal L_{\mathrm{ref}}(\omega)\bm G_{\mathrm{ref}}(\bm{x},\bm{x}';\omega)
	=\bm I\,\delta(\bm{x}-\bm{x}').
	\label{eq:green_ref}
\end{equation}
For a source current density $\bm J_{s,e}$, the reference incident field is
\begin{equation}
	\bm E_{\mathrm{inc},e}^{\mathrm{ref}}(\bm{x},\omega_e)
	=-j\omega_e\mu_0\int
	\bm G_{\mathrm{ref}}(\bm{x},\bm{x}';\omega_e)
	\bm J_{s,e}(\bm{x}',\omega_e)\,\mathrm d\bm{x}'.
	\label{eq:incident_field}
\end{equation}
If $\mathcal W_{m,e}$ denotes the frozen linear receive functional for receiver $m$ and event $e$, the effective reference measurement kernel can be written as
\begin{equation}
	\begin{aligned}
		K_{me}^{(0)}(\bm{x};\omega_e)
		\coloneqq{} & \omega_e^2\mu_0\varepsilon_{\mathrm{ref}}(\omega_e,\bm{x})
		\\[-1mm]
		            & \quad\times\mathcal W_{m,e}\!\left[
			\bm G_{\mathrm{ref}}(\cdot,\bm{x};\omega_e)
			\bm E_{\mathrm{inc},e}^{\mathrm{ref}}(\bm{x},\omega_e)
			\right].
	\end{aligned}
	\label{eq:effective_reference_measurement_kernel}
\end{equation}
This kernel is fixed after the reference state, geometry, event frequencies, and receive processing have been selected. Cross-polarized coupling and instrument-specific effects not represented by this kernel remain in the modeling remainder or instrumental-disturbance term.

The exact constitutive increment is
\begin{equation}
	\Delta\varepsilon_{c,e}(\bm{x}) = \mathcal F_{\mathrm{CC}}(\omega_e,\bm{\mu}_{\mathrm{ref}}+\delta\bm{\mu})-\mathcal F_{\mathrm{CC}}(\omega_e,\bm{\mu}_{\mathrm{ref}}).
	\label{eq:delta_eps_exact}
\end{equation}
The corresponding exact reference-normalized contrast is
\begin{equation}
	\xi_e^{\mathrm{ex}}(\bm{x};\delta\bm{\mu})
	=\frac{\Delta\varepsilon_{c,e}(\bm{x})}
	{\varepsilon_{\mathrm{ref}}(\omega_e,\bm{x})}.
	\label{eq:xi_def}
\end{equation}
To measure local perturbation size without mixing the units of the five Cole--Cole parameters, introduce a fixed positive diagonal scale matrix $\bm S_{\mu}$ and define
\begin{equation}
	\delta\bm u(\bm{x})\coloneqq\bm S_{\mu}^{-1}\delta\bm{\mu}(\bm{x}).
	\label{eq:scaled_parameter_residual}
\end{equation}
At the reference state, the constitutive Jacobian gives
\begin{equation}
	\mathbf J_{\mathrm{CC},e}(\bm{x}) = \left.\frac{\partial\mathcal F_{\mathrm{CC}}(\omega_e,\bm{\mu})}{\partial\bm{\mu}}\right|_{\bm{\mu}=\bm{\mu}_{\mathrm{ref}}},
	\label{eq:cole_cole_jacobian}
\end{equation}
The first-order complex-permittivity increment and its normalized form are
\begin{equation}
	\delta\varepsilon_{c,e}^{(1)}(\bm{x})
	=\bm J_{\mathrm{CC},e}(\bm{x})\delta\bm{\mu}(\bm{x}),
	\label{eq:delta_eps_first_order}
\end{equation}
\begin{equation}
	\bm b_e^{\mathrm T}(\bm{x})\coloneqq
	\frac{\bm J_{\mathrm{CC},e}(\bm{x})}
	{\varepsilon_{\mathrm{ref}}(\omega_e,\bm{x})}.
	\label{eq:normalized_constitutive_jacobian}
\end{equation}
and the first-order normalized contrast can be written as
\begin{equation}
	\xi_e^{(1)}(\bm{x})=\bm b_e(\bm{x})^{\mathrm T}\delta\bm{\mu}(\bm{x}),
	\label{eq:xi_first_order}
\end{equation}
where $\bm b_e$ absorbs the constitutive Jacobian and the selected normalization. The constitutive remainder is
\begin{equation}
	\rho_{\mathrm{CC},e}(\bm{x}) = \xi_e(\bm{x})-\xi_e^{(1)}(\bm{x}).
	\label{eq:constitutive_remainder}
\end{equation}
Under local differentiability of the constitutive map,
\begin{equation}
	\frac{\lVert \rho_{\mathrm{CC},e}(\delta\bm{\mu})\rVert}
	{\lVert\delta\bm u\rVert}
	\longrightarrow 0
	\qquad\text{as}\qquad
	\lVert\delta\bm u\rVert\longrightarrow0.
	\label{eq:constitutive_remainder_local}
\end{equation}
It is locally second order under the usual differentiability conditions, but no universal small-remainder condition is imposed for the complete electromagnetic problem.

Let $\mathscr F_{me}[\bm{\mu}]$ denote the complete noise-free electromagnetic measurement at receiver $m$ under event $e$. The exact reference-state incremental observation is
\begin{equation}
	y_{me}^{\mathrm{ex}}
	\coloneqq
	\mathscr F_{me}[\bm{\mu}_{\mathrm{ref}}+\delta\bm{\mu}]
	-\mathscr F_{me}[\bm{\mu}_{\mathrm{ref}}].
	\label{eq:exact_incremental_observation}
\end{equation}

The receiver-level first-order response and its complete model remainder can be written as
\begin{equation}
	\begin{aligned}
		y_{me}^{(1)}           & \coloneqq\int_D K_{me}^{(0)}(\bm{x};\omega_e)\xi_e^{(1)}(\bm{x})\,\mathrm d\bm{x}, \\
		r_{me}(\delta\bm{\mu}) & \coloneqq y_{me}^{\mathrm{ex}}-y_{me}^{(1)}.
	\end{aligned}
	\label{eq:event_model_remainder}
\end{equation}

The fixed reference kernel maps the first-order contrast to an event response:
\begin{equation}
	y_{e}^{(1)} = \int_{\Omega}\mathbf K_e^{(0)}(\bm{x})\xi_e^{(1)}(\bm{x})\,\mathrm d\bm{x}.
	\label{eq:first_order_event_response}
\end{equation}
The observation model is
\begin{equation}
	y_e=y_e^{(1)}+r_e+n_e,
	\label{eq:echo_channel}
\end{equation}
where $r_e$ collects higher-order constitutive terms, reference-kernel changes, multiple scattering, accumulated phase distortion, interface feedback, target--background coupling, and instrument or modeling disturbances not represented by $n_e$.

Physically, the DBA term represents the field generated by a single incremental interaction evaluated at the selected reference state. The reference Green tensor and incident field remain fixed, and the constitutive perturbation is mapped to the receiver through the reference measurement kernel. The resulting first-order term describes a local tangent of the propagation map around the selected working state.

Thus, \eqref{eq:echo_channel} is used as an operational first-order DBA model. Its quantitative validity is assessed in both the observation and task domains. \cref{sec:experiments} evaluates this validity with $D_{\mathrm{obs}}$ and $D_A$ over the stated mismatch conditions.

\subsection{Discrete single-snapshot representation}\label{subsec:vectorization}

Discretize the region into $P$ voxels. Let $\delta\bm\mu$, $\delta\bm\mu_{\mathrm{sh}}$, and $\delta\bm\mu_{\mathrm{an}}$ denote the corresponding stacked parameter vectors.
Let $\bm{x}_p$ denote the center of voxel $p$ and let $\Delta_p$ be its integration weight. The parameter residual and its two components are stacked voxel by voxel as
\begin{equation}
	\delta\underline{\bm{\mu}}
	\coloneqq
	\left[\delta\bm{\mu}^{\mathrm T}(\bm{x}_1),\ldots,
		\delta\bm{\mu}^{\mathrm T}(\bm{x}_P)\right]^{\mathrm T}
	\in\mathbb R^{5P},
	\label{eq:delta_mu_stack}
\end{equation}
\begin{equation}
	\delta\underline{\bm{\mu}}_{\mathrm{sh}}
	\coloneqq
	\left[\delta\bm{\mu}_{\mathrm{sh}}^{\mathrm T}(\bm{x}_1),\ldots,
	\delta\bm{\mu}_{\mathrm{sh}}^{\mathrm T}(\bm{x}_P)\right]^{\mathrm T},
	\label{eq:delta_mu_sh_stack}
\end{equation}
\begin{equation}
	\delta\underline{\bm{\mu}}_{\mathrm{an}}
	\coloneqq
	\left[\delta\bm{\mu}_{\mathrm{an}}^{\mathrm T}(\bm{x}_1),\ldots,
	\delta\bm{\mu}_{\mathrm{an}}^{\mathrm T}(\bm{x}_P)\right]^{\mathrm T}.
	\label{eq:delta_mu_an_stack}
\end{equation}
\begin{equation}
	\delta\underline{\bm{\mu}}
	=\delta\underline{\bm{\mu}}_{\mathrm{sh}}
	+\delta\underline{\bm{\mu}}_{\mathrm{an}}.
	\label{eq:delta_mu_stack_split}
\end{equation}

For event $e$, define a constitutive map $\mathbf B_e$, a propagation matrix $\mathbf H_e$, and the event tangent operator
\begin{equation}
	\mathbf B_e\in\mathbb C^{P\times5P}.
	\label{eq:Be_dimension}
\end{equation}
Its nonzero entries are
\begin{equation}
	\left[\mathbf B_e\right]_{p,5(p-1)+q}
	=\left[\bm b_e(\bm{x}_p)\right]_q,
	\qquad q=1,\ldots,5.
	\label{eq:Be_entry}
\end{equation}
Here $q$ indexes the five entries of the fixed Cole--Cole vector in \eqref{eq:mu_def}. A different constitutive parameterization would replace $5$ by its parameter dimension, but the present manuscript uses the five-parameter Cole--Cole model throughout.
The discretized first-order contrast is
\begin{equation}
	\bm\xi_e^{(1)}=\mathbf B_e\delta\underline{\bm{\mu}}\in\mathbb C^P.
	\label{eq:xi_stack}
\end{equation}
The reference propagation matrix satisfies
\begin{equation}
	\mathbf H_e\in\mathbb C^{N_r\times P},
	\label{eq:He_dimension}
\end{equation}
with entries
\begin{equation}
	\left[\mathbf H_e\right]_{m,p}
	=K_{me}^{(0)}(\bm{x}_p;\omega_e)\Delta_p.
	\label{eq:He_entry}
\end{equation}
\begin{equation}
	\mathbf L_e=\mathbf H_e\mathbf B_e.
	\label{eq:Le_def}
\end{equation}
For receiver-fastest stacking within each event, define
\begin{equation}
	\bm y_e\coloneqq[y_{1e},\ldots,y_{N_r e}]^{\mathrm T}.
	\label{eq:y_stack}
\end{equation}
\begin{equation}
	\bm r_e\coloneqq[r_{1e},\ldots,r_{N_r e}]^{\mathrm T}.
	\label{eq:r_stack}
\end{equation}
\begin{equation}
	\bm n_e\coloneqq[n_{1e},\ldots,n_{N_r e}]^{\mathrm T}.
	\label{eq:n_stack}
\end{equation}
The event model becomes
\begin{equation}
	y_e=\mathbf L_e\delta\bm\mu+r_e+n_e
	=\mathbf L_e\delta\bm\mu_{\mathrm{sh}}+\mathbf L_e\delta\bm\mu_{\mathrm{an}}+r_e+n_e.
	\label{eq:discrete_channel_model}
\end{equation}
Equivalently, the first-order event response separates as
\begin{equation}
	\bm y_e=\mathbf L_e\delta\underline{\bm{\mu}}_{\mathrm{sh}}
	+\mathbf L_e\delta\underline{\bm{\mu}}_{\mathrm{an}}
	+\bm r_e+\bm n_e.
	\label{eq:event_background_anomaly_split}
\end{equation}
The shared and anomaly components are
\begin{equation}
	\bm c_e^{(\mathrm{sh})}\coloneqq\mathbf L_e\delta\underline{\bm{\mu}}_{\mathrm{sh}}.
	\label{eq:event_shared_response}
\end{equation}
\begin{equation}
	\bm c_e^{(\mathrm{an})}\coloneqq\mathbf L_e\delta\underline{\bm{\mu}}_{\mathrm{an}}.
	\label{eq:event_anomaly_response}
\end{equation}
The joint observation uses receiver-fastest and event-slowest stacking,
\begin{equation}
	\bm y=\left[y_1^{\mathrm T},\ldots,y_{N_e}^{\mathrm T}\right]^{\mathrm T},\qquad
	\mathbf L=\left[\mathbf L_1^{\mathrm T},\ldots,\mathbf L_{N_e}^{\mathrm T}\right]^{\mathrm T},
	\label{eq:joint_snapshot}
\end{equation}
The joint tangent, remainder, and noise vectors are
\begin{equation}
	\mathbf L=\left[\mathbf L_1^{\mathrm T},\ldots,\mathbf L_{N_e}^{\mathrm T}\right]^{\mathrm T},
	\label{eq:joint_tangent_operator}
\end{equation}
\begin{equation}
	\bm r=\left[\bm r_1^{\mathrm T},\ldots,\bm r_{N_e}^{\mathrm T}\right]^{\mathrm T}.
	\label{eq:joint_remainder}
\end{equation}
\begin{equation}
	\bm n=\left[\bm n_1^{\mathrm T},\ldots,\bm n_{N_e}^{\mathrm T}\right]^{\mathrm T}.
	\label{eq:joint_noise}
\end{equation}
and therefore satisfies
\begin{equation}
	\bm y=\mathbf L\delta\bm\mu_{\mathrm{sh}}+\mathbf L\delta\bm\mu_{\mathrm{an}}+\bm r+\bm n.
	\label{eq:joint_single_snapshot_model}
\end{equation}
The joint first-order shared response is
\begin{equation}
	\bm c^{(\mathrm{sh})}\coloneqq\mathbf L\delta\underline{\bm{\mu}}_{\mathrm{sh}}.
	\label{eq:joint_shared_response}
\end{equation}
The single snapshot is one realization of the complete $N_eN_r$-dimensional vector. An offline ensemble used to estimate residual moments describes prior or simulated uncertainty and does not create additional temporal snapshots.

\section{Reference--Background Residual in the Joint Observation Domain}\label{sec:observation_statistics}\label{sec:observation}

\subsection{Exact background difference and first-order response}\label{subsec:exact_first_order}\label{subsec:ob}

For an anomaly-free physical background, let the vector-valued event forward operator be
\begin{equation}
	\mathscr F_e[\bm\mu]
	\coloneqq
	\left[\mathscr F_{1e}[\bm\mu],\ldots,\mathscr F_{N_r e}[\bm\mu]\right]^{\mathrm T}.
	\label{eq:event_forward_operator}
\end{equation}
The exact background-state difference is
\begin{equation}
	c_e^{(\mathrm{sh},\mathrm{ex})}=\mathscr F_e[\bm\mu_b]-\mathscr F_e[\bm\mu_{\mathrm{ref}}].
	\label{eq:exact_shared_background_response}
\end{equation}
The first-order shared response is
\begin{equation}
	c_e^{(\mathrm{sh})}=\mathbf L_e\delta\bm\mu_{\mathrm{sh}},
	\label{eq:ob_event_shared_response}
\end{equation}
and the joint first-order response is
\begin{equation}
	\bm c^{(\mathrm{sh})}=\mathbf L\delta\underline{\bm{\mu}}_{\mathrm{sh}}.
	\label{eq:ob_joint_shared_response}
\end{equation}
and the background-specific remainder is defined by
\begin{equation}
	r_e^{(\mathrm{sh})}=c_e^{(\mathrm{sh},\mathrm{ex})}-c_e^{(\mathrm{sh})}.
	\label{eq:shared_background_remainder}
\end{equation}
The eventwise exact-to-first-order relation is equivalently
\begin{equation}
	c_e^{(\mathrm{sh},\mathrm{ex})}
	=\mathbf L_e\delta\underline{\bm{\mu}}_{\mathrm{sh}}+r_e^{(\mathrm{sh})}.
	\label{eq:exact_first_order_shared_relation}
\end{equation}
The exact event responses and their remainders are stacked as
\begin{equation}
	\bm c^{(\mathrm{sh,ex})}
	=\left[(c_1^{(\mathrm{sh,ex})})^{\mathrm T},\ldots,
	(c_{N_e}^{(\mathrm{sh,ex})})^{\mathrm T}\right]^{\mathrm T},
	\label{eq:joint_exact_shared_response}
\end{equation}
\begin{equation}
	\bm r^{(\mathrm{sh})}
	=\left[(r_1^{(\mathrm{sh})})^{\mathrm T},\ldots,
	(r_{N_e}^{(\mathrm{sh})})^{\mathrm T}\right]^{\mathrm T}.
	\label{eq:joint_shared_remainder}
\end{equation}
Thus,
\begin{equation}
	c_e^{(\mathrm{sh},\mathrm{ex})}=c_e^{(\mathrm{sh})}+r_e^{(\mathrm{sh})}.
	\label{eq:joint_exact_first_order_relation}
\end{equation}
Closed-form covariance expressions below describe $c^{(\mathrm{sh})}$. Full-wave differences and $r^{(\mathrm{sh})}$ are used to determine whether that first-order quantity remains operationally interpretable.

\subsection{FDA frequency states and assignment-dependent propagation}
\label{subsec:local_fda_coding}

Define a finite set of $K_f$ frequency states by
\begin{equation}
	\mathcal W=\left\{\omega_1^{(f)},\ldots,\omega_{K_f}^{(f)}\right\},
	\qquad 1\leq K_f\leq N_e.
	\label{eq:frequency_state_set}
\end{equation}
The one-hot assignment of these states to the physical transmit events is represented by
\begin{equation}
	\mathbf Z\in\{0,1\}^{N_e\times K_f},
	\label{eq:assignment_matrix_set}
\end{equation}
subject to
\begin{equation}
	\sum_{k=1}^{K_f}Z_{ek}=1,
	\qquad e=1,\ldots,N_e.
	\label{eq:assignment_row_constraint}
\end{equation}
Let $a_{\mathbf Z}(e)$ be the selected state index, so that
\begin{equation}
	Z_{e,a_{\mathbf Z}(e)}=1,
	\label{eq:assignment_index}
\end{equation}
and
\begin{equation}
	\omega_e(\mathbf Z)=\omega_{a_{\mathbf Z}(e)}^{(f)}.
	\label{eq:event_assigned_frequency}
\end{equation}
The matrix $\mathbf Z$ represents frequency-state assignment only. Amplitude, phase, waveform, and general coding-matrix variations lie outside the assignment family studied here.

The assigned event frequency is $\omega_e(\mathbf Z)$, and the event operator becomes
\begin{equation}
	\mathbf L_e(\mathbf Z)=\mathbf H_e\bigl(\omega_e(\mathbf Z)\bigr)\mathbf B_e\bigl(\omega_e(\mathbf Z)\bigr),
	\label{eq:assignment_event_operator}
\end{equation}
with joint operator
\begin{equation}
	\mathbf L(\mathbf Z)=\left[\mathbf L_1(\mathbf Z)^{\mathrm T},\ldots,
		\mathbf L_{N_e}(\mathbf Z)^{\mathrm T}\right]^{\mathrm T}.
	\label{eq:assignment_joint_operator}
\end{equation}
For the assigned state, the constitutive and propagation factors can also be written as
\begin{equation}
	\mathbf B_e(\mathbf Z)\coloneqq\mathbf B\!\left(\omega_{a_{\mathbf Z}(e)}^{(f)}\right),
	\label{eq:assignment_constitutive_matrix}
\end{equation}
\begin{equation}
	\mathbf H_e(\mathbf Z)\coloneqq\mathbf H\!\left(\bm s_e,\omega_{a_{\mathbf Z}(e)}^{(f)}\right).
	\label{eq:assignment_propagation_matrix}
\end{equation}
The shared response and its covariance are consequently
\begin{equation}
	c^{(\mathrm{sh})}(\mathbf Z)=\mathbf L(\mathbf Z)\delta\bm\mu_{\mathrm{sh}},
	\qquad
	\mathbf R_{\mathrm{sh}}(\mathbf Z)=\mathbf L(\mathbf Z)\mathbf C_\mu\mathbf L(\mathbf Z)^{\mathrm H}.
	\label{eq:fda_modulated_covariance}
\end{equation}
The response component in this relation is explicitly
\begin{equation}
	\bm c^{(\mathrm{sh})}(\mathbf Z)
	=\mathbf L(\mathbf Z)\delta\underline{\bm{\mu}}_{\mathrm{sh}}.
	\label{eq:assignment_shared_response}
\end{equation}
For a uniform linear FDA, $K_f=N_e$ and each state is used once in monotone order. Its assignment matrix and frequency states satisfy
\begin{equation}
	\left[\mathbf Z_{\mathrm{lin}}\right]_{ek}
	=\begin{cases}1, & e=k, \\0,&e\neq k,
	\end{cases}
	\label{eq:linear_fda_assignment}
\end{equation}
\begin{equation}
	\omega_k^{(f)}=\omega_0+(k-1)\Delta\omega,
	\qquad k=1,\ldots,N_e.
	\label{eq:linear_fda_frequency_set}
\end{equation}
The zero-frequency-diversity case is
\begin{equation}
	K_f=1,
	\label{eq:single_frequency_state}
\end{equation}
and its eventwise response remains
\begin{equation}
	\bm c_e^{(\mathrm{sh})}
	=\mathbf L\!\left(\bm s_e,\omega_1^{(f)}\right)
	\delta\underline{\bm{\mu}}_{\mathrm{sh}}.
	\label{eq:zero_diversity_event_response}
\end{equation}
The linear FDA special case assigns frequencies monotonically with a transmitter-dependent step. The zero-diversity case has $K_f=1$ and still permits nonzero event covariance because event geometry and propagation kernels can differ. The assignment family studied here contains frequency-state assignments only. Equation~\eqref{eq:fda_modulated_covariance} makes the causal separation explicit: $\mathbf C_\mu$ describes the common residual source, while $\mathbf L(\mathbf Z)$ describes FDA-dependent propagation organization.

\subsection{Residual statistics and covariance blocks}\label{subsec:covariance}

The physical residual ensemble is decomposed as
\begin{equation}
	\delta\underline{\bm{\mu}}_{\mathrm{sh}}=\overline{\bm d}_{\mathrm{sh}}+\bm\eta_{\mathrm{sh}},
	\qquad
	\mathbb E[\bm\eta_{\mathrm{sh}}]=\bm 0,
	\qquad
	\mathbf C_\mu=\mathbb E\left[\bm\eta_{\mathrm{sh}}\bm\eta_{\mathrm{sh}}^{\mathrm T}\right].
	\label{eq:shared_parameter_decomposition}
\end{equation}
The mean and centered-fluctuation conditions can be recorded separately as
\begin{equation}
	\overline{\bm d}_{\mathrm{sh}}
	\coloneqq\mathbb E\left[\delta\underline{\bm{\mu}}_{\mathrm{sh}}\right].
	\label{eq:shared_parameter_mean}
\end{equation}
\begin{equation}
	\mathbb E[\bm\eta_{\mathrm{sh}}]=\bm 0.
	\label{eq:shared_fluctuation_zero_mean}
\end{equation}
The parameter-domain covariance is
\begin{equation}
	\mathbf C_\mu\coloneqq
	\mathbb E\left[\bm\eta_{\mathrm{sh}}\bm\eta_{\mathrm{sh}}^{\mathrm T}\right].
	\label{eq:parameter_covariance}
\end{equation}
The covariance may contain spatial correlation and cross-correlation among the five constitutive parameters. The continuous parameter-domain covariance is specified by the following cross-covariance kernel and its discretization. For the continuous residual field, let $\eta_{\mathrm{sh},a}(\bm{x})$ denote parameter component $a$.
\begin{equation}
	K_{ab}(\bm{x},\bm{x}')
	\coloneqq\operatorname{Cov}\left[\eta_{\mathrm{sh},a}(\bm{x}),\eta_{\mathrm{sh},b}(\bm{x}')\right],
	\qquad a,b=1,\ldots,5,
	\label{eq:parameter_cross_covariance_kernel}
\end{equation}
\begin{equation}
	\left[\mathbf C_\mu\right]_{5(p-1)+a,5(p'-1)+b}
	=K_{ab}(\bm{x}_p,\bm{x}_{p'}).
	\label{eq:continuous_discrete_covariance_relation}
\end{equation}
Substitution of the residual decomposition into the shared response gives
\begin{equation}
	\bm c^{(\mathrm{sh})}(\mathbf Z)
	=\mathbf L(\mathbf Z)\overline{\bm d}_{\mathrm{sh}}
	+\mathbf L(\mathbf Z)\bm\eta_{\mathrm{sh}}.
	\label{eq:shared_response_mean_random_split}
\end{equation}
\begin{equation}
	\bm m_{\mathrm{sh}}(\mathbf Z)=\mathbf L(\mathbf Z)\overline{\bm d}_{\mathrm{sh}},
	\label{eq:msh_def}
\end{equation}
\begin{equation}
	\mathbf R_{\mathrm{sh}}(\mathbf Z)=\mathbf L(\mathbf Z)\mathbf C_\mu\mathbf L(\mathbf Z)^{\mathrm H}.
	\label{eq:Rsh_def}
\end{equation}
The covariance has the closed form
\begin{equation}
	\mathbf R_{\mathrm{sh}}(\mathbf Z)
	=\mathbf L(\mathbf Z)\mathbf C_\mu\mathbf L^{\mathrm H}(\mathbf Z).
	\label{eq:observation_covariance_closed_form}
\end{equation}
For event blocks $e$ and $e'$,
\begin{equation}
	\mathbf R_{\mathrm{sh}}^{(e,e')}(\mathbf Z)=
	\mathbf L_e(\mathbf Z)\mathbf C_\mu\mathbf L_{e'}(\mathbf Z)^{\mathrm H}.
	\label{eq:Rsh_block_def}
\end{equation}
The block matrix and eventwise mean can be written as
\begin{equation}
	\mathbf R_{\mathrm{sh}}(\mathbf Z)
	=\left[\mathbf R_{\mathrm{sh}}^{(e,e')}(\mathbf Z)\right]_{e,e'=1}^{N_e},
	\label{eq:Rsh_block}
\end{equation}
\begin{equation}
	\bm m_e(\mathbf Z)=\mathbf L_e(\mathbf Z)\overline{\bm d}_{\mathrm{sh}}.
	\label{eq:mn_def}
\end{equation}
An off-event block can be nonzero whenever the two event operators observe correlated parts of the common residual. Different frequency states are required for a cross-frequency interpretation, but frequency separation neither creates nor guarantees dependence.
An off-event block is a cross-frequency block only if
\begin{equation}
	\omega_{a_{\mathbf Z}(e)}^{(f)}
	\neq
	\omega_{a_{\mathbf Z}(e')}^{(f)}.
	\label{eq:cross_frequency_block_condition}
\end{equation}
The condition for a nonzero off-event block is
\begin{equation}
	\mathbf L_e(\mathbf Z)\mathbf C_\mu\mathbf L_{e'}^{\mathrm H}(\mathbf Z)\neq\bm 0.
	\label{eq:nonzero_off_event_condition}
\end{equation}

Three covariance models are used to describe controlled approximations:
\begin{align}
	\mathbf R_{\mathrm{full}} & = \left[\mathbf R^{(e,e')}_{\mathrm{sh}}\right]_{e,e'=1}^{N_e},
	\label{eq:full_covariance_model}                                                                                                             \\
	\mathbf R_{\mathrm{blk}}  & = \operatorname{blkdiag}\left(\mathbf R^{(1,1)}_{\mathrm{sh}},\ldots,\mathbf R^{(N_e,N_e)}_{\mathrm{sh}}\right),
	\label{eq:event_block_covariance_model}                                                                                                      \\
	\mathbf R_{\mathrm{diag}} & = \operatorname{diag}\left(\operatorname{diag}\mathbf R_{\mathrm{blk}}\right).
	\label{eq:diagonal_covariance_model}
\end{align}
The event-block model removes off-event covariance but retains within-event receiver covariance. The diagonal model also removes receiver dependence within each event. Both are approximations to the full covariance, not alternative physical sources.

\section{Transfer to Fixed Linear Task Domains}\label{sec:task_transfer}

\subsection{Fixed task map and task-domain moments}\label{subsec:fixed_linear_task}\label{subsec:task_statistics}

Let $\mathbf A$ be a fixed linear task operator selected from the reference state, event configuration, task definition, and regularization parameters before a residual realization is observed. For explicit dimensions, the fixed task map is
\begin{equation}
	\mathbf A\in\mathbb C^{P_A\times N_eN_r}.
	\label{eq:fixed_task_operator_dimension}
\end{equation}
Its output is
\begin{equation}
	\bm z_A\coloneqq\mathbf A\bm y.
	\label{eq:fixed_linear_task_map}
\end{equation}
The adjective fixed means that the reference state, event configuration, task definition, and operator hyperparameters are selected before the current residual, anomaly, or noise realization. Linearity requires
\begin{equation}
	\mathbf A(a\bm y_1+b\bm y_2)=a\mathbf A\bm y_1+b\mathbf A\bm y_2,
	\qquad a,b\in\mathbb C.
	\label{eq:task_operator_linearity}
\end{equation}
The first-order anomaly response and the input decomposition are
\begin{equation}
	\bm c^{(\mathrm{an})}\coloneqq\mathbf L\delta\underline{\bm{\mu}}_{\mathrm{an}},
	\label{eq:joint_anomaly_response}
\end{equation}
\begin{equation}
	\bm y=\bm c^{(\mathrm{an})}+\bm c^{(\mathrm{sh})}+\bm r+\bm n.
	\label{eq:task_input_decomposition}
\end{equation}
The four task components are defined by
\begin{equation}
	\bm z_A^{(\mathrm{an})}\coloneqq\mathbf A\bm c^{(\mathrm{an})},
	\label{eq:task_anomaly_component}
\end{equation}
\begin{equation}
	\bm e_A^{(\mathrm{sh})}\coloneqq\mathbf A\bm c^{(\mathrm{sh})},
	\label{eq:task_shared_component}
\end{equation}
\begin{equation}
	\bm e_A^{(r)}\coloneqq\mathbf A\bm r.
	\label{eq:task_remainder_component}
\end{equation}
\begin{equation}
	\bm e_A^{(n)}\coloneqq\mathbf A\bm n.
	\label{eq:task_noise_component}
\end{equation}
The shared-residual task component is not tied to a particular inversion algorithm. In a spatial task domain it can appear as a structured pseudo-anomaly, whereas in a general task domain it denotes the output driven by the reference--background residual. The exact background-state task component and its first-order relation are
\begin{equation}
	\bm e_A^{(\mathrm{sh,ex})}\coloneqq\mathbf A\bm c^{(\mathrm{sh,ex})},
	\label{eq:task_exact_shared_component}
\end{equation}
\begin{equation}
	\bm e_A^{(\mathrm{sh,ex})}=\bm e_A^{(\mathrm{sh})}+\mathbf A\bm r^{(\mathrm{sh})}.
	\label{eq:task_exact_first_order_relation}
\end{equation}
\begin{equation}
	\bm z_A=\bm z_A^{(\mathrm{an})}+\bm e_A^{(\mathrm{sh})}+\bm e_A^{(r)}+\bm e_A^{(n)}.
	\label{eq:task_output_decomposition}
\end{equation}
The shared task mean, covariance, and mean-square risk are
\begin{align}
	\bm m_{A,\mathrm{sh}}     & = \mathbf A\bm m_{\mathrm{sh}},
	\label{eq:task_shared_mean}                                                                 \\
	\mathbf R_{A,\mathrm{sh}} & = \mathbf A\mathbf R_{\mathrm{sh}}\mathbf A^{\mathrm H},
	\label{eq:task_shared_covariance}                                                           \\
	J_A^{(\mathrm{sh})}       & = \mathbb E\left[\lVert\bm e_A^{(\mathrm{sh})}\rVert_2^2\right]
	=\lVert\mathbf A\bm m_{\mathrm{sh}}\rVert_2^2+\operatorname{tr}\left(\mathbf A\mathbf R_{\mathrm{sh}}\mathbf A^{\mathrm H}\right).
	\label{eq:task_shared_risk}
\end{align}

\begin{proposition}[Task-domain moment and risk transfer]\label{prop:task_moment_transfer}
	For any fixed linear task operator and finite second moments, the task mean, covariance, and risk are given by \eqref{eq:task_shared_mean}--\eqref{eq:task_shared_risk}. Linearity gives $\mathbb E[\mathbf A\bm c]=\mathbf A\mathbb E[\bm c]$ and $\operatorname{Cov}(\mathbf A\bm c)=\mathbf A\operatorname{Cov}(\bm c)\mathbf A^{\mathrm H}$. The squared-norm identity separates the mean and covariance contributions. \cref{sec:experiments} evaluates these identities through its existing predicted-risk and empirical-energy calibration.
\end{proposition}
\begin{proof}
	Linearity gives $\mathbb E[\mathbf A\bm c]=\mathbf A\mathbb E[\bm c]$ and $\operatorname{Cov}(\mathbf A\bm c)=\mathbf A\operatorname{Cov}(\bm c)\mathbf A^{\mathrm H}$. Expanding the squared norm about the mean gives the two terms in $J_A^{(\mathrm{sh})}$.
\end{proof}

The centered task component and the mean--covariance energy identity are
\begin{equation}
	\bm e_A^{(\mathrm{sh})}-\bm m_{A,\mathrm{sh}}
	=\mathbf A\left(\bm c^{(\mathrm{sh})}-\bm m_{\mathrm{sh}}\right).
	\label{eq:centered_task_shared_component}
\end{equation}
\begin{equation}
	\mathbb E\left[\lVert\bm x\rVert_2^2\right]
	=\lVert\mathbb E[\bm x]\rVert_2^2+\operatorname{tr}\left(\operatorname{Cov}[\bm x]\right).
	\label{eq:mean_covariance_energy_identity}
\end{equation}
The risk separates systematic mean bias from random host variability. Its two contributions are
\begin{equation}
	J_{A,\mathrm{mean}}^{(\mathrm{sh})}
	\coloneqq\lVert\mathbf A\bm m_{\mathrm{sh}}\rVert_2^2.
	\label{eq:task_mean_risk}
\end{equation}
\begin{equation}
	J_{A,\mathrm{cov}}^{(\mathrm{sh})}
	\coloneqq\operatorname{tr}\left(\mathbf A\mathbf R_{\mathrm{sh}}\mathbf A^{\mathrm H}\right).
	\label{eq:task_covariance_risk}
\end{equation}
This is a task-dependent measure of residual energy. Image reliability and target detection require separate task-specific metrics.

\subsection{Projected covariance and first-order task discrepancy}\label{subsec:task_covariance_discrepancy}

For $\star\in\{\mathrm{full},\mathrm{blk},\mathrm{diag}\}$, the covariance model used in the observation domain is transferred to the task domain as follows.
\begin{proposition}[Projected covariance identity]\label{prop:task_covariance_projection}
	\begin{equation}
		\mathbf R_{A,\star}^{(\mathrm{sh})}=\mathbf A\mathbf R_{\star}\mathbf A^{\mathrm H},
		\qquad \star\in\{\mathrm{full},\mathrm{blk},\mathrm{diag}\}.
		\label{eq:task_covariance_model_projection}
	\end{equation}
	Consequently, the difference between the full model and any reduced model is
	\begin{equation}
		\Delta\mathbf R_{A,\star}^{(\mathrm{sh})}
		=\mathbf A(\mathbf R_{\mathrm{full}}-\mathbf R_{\star})\mathbf A^{\mathrm H},
		\qquad \star\in\{\mathrm{blk},\mathrm{diag}\}.
		\label{eq:task_covariance_model_difference}
	\end{equation}
\end{proposition}
\begin{proof}
	Substitution of the observation-domain covariance model into the linear task map gives the first identity. Subtracting the projected reduced model from the projected full model gives the second identity.
\end{proof}

The absolute and relative projected errors used for the two approximations are
\begin{equation}
	E_{A,\star}^{\mathrm{abs}}
	\coloneqq\left\lVert\mathbf R_{A,\mathrm{full}}^{(\mathrm{sh})}-\mathbf R_{A,\star}^{(\mathrm{sh})}\right\rVert_{\mathrm F},
	\qquad \star\in\{\mathrm{blk},\mathrm{diag}\},
	\label{eq:absolute_task_covariance_error}
\end{equation}
\begin{equation}
	E_{A,\star}^{\mathrm{rel}}
	\coloneqq
	\frac{E_{A,\star}^{\mathrm{abs}}}{\left\lVert\mathbf R_{A,\mathrm{full}}^{(\mathrm{sh})}\right\rVert_{\mathrm F}}.
	\label{eq:relative_task_covariance_error}
\end{equation}
\cref{sec:experiments} reports these projected errors, together with the full projected covariance norms and operator norms, for the three fixed task ports.

The exact and first-order shared-residual task components satisfy
\begin{proposition}[First-order task discrepancy transfer]\label{prop:task_discrepancy_transfer}
	For a fixed linear task map, the exact-to-first-order discrepancy is the projected remainder in \eqref{eq:task_exact_first_order_relation}.
\end{proposition}
\begin{proof}
	The identity follows by applying the linear task operator to the exact-to-first-order observation relation.
\end{proof}

The mismatch sweep in \cref{sec:experiments} uses $D_{\mathrm{obs}}$ and $D_A$ to delimit the conditions under which the first-order task interpretation remains quantitatively useful. This transfer relation is a discrepancy statement for the fixed task map; it does not provide a nonlinear forward or inverse stability result.

\subsection{Task-risk interface for finite reference screening}\label{subsec:finite_reference_task_interface}

Since the reference state determines both the task operator and the shared-residual moments, its choice directly affects the task-domain output and risk. Under the fixed-task formulation above, each admissible reference state $r$ determines a corresponding operator $\mathbf A_r$ and shared-residual moments $(\bm m_{\mathrm{sh},r},\mathbf R_{\mathrm{sh},r})$. A finite library can therefore be screened by comparing task risks, which feed back into reference-state selection.

For a finite admissible reference library $\mathcal R$, the task-risk interface evaluates
\begin{equation}
	\mathcal R_{\mathrm{adm}}\subseteq\mathcal R.
	\label{eq:admissible_reference_library}
\end{equation}
\begin{equation}
	J_A(r)=\left\lVert\mathbf A_r\bm m_{\mathrm{sh},r}\right\rVert_2^2+
	\operatorname{tr}\left(\mathbf A_r\mathbf R_{\mathrm{sh},r}\mathbf A_r^{\mathrm H}\right),
	\qquad r\in\mathcal R,
	\label{eq:finite_reference_task_risk}
\end{equation}
and selects
\begin{equation}
	r^\star\in\arg\min_{r\in\mathcal R}J_A(r).
	\label{eq:finite_reference_selection_rule}
\end{equation}
The moments are estimated on an independent training ensemble, candidates are selected on independent validation realizations, and the selected candidate is assessed on independent test realizations. The interface defines a finite reference-state ranking problem for a fixed task, admissible library, and first-order validity domain.

\section{Numerical Experiments and Evidence Boundaries}\label{sec:experiments}

\subsection{Common configuration and residual ensemble}\label{subsec:exp_config}

The analytical and full-wave experiments used the same event semantics and receiver-fastest/event-slowest stacking rule. The $48$-dimensional observation used six transmit events and eight receive channels. The reference frequency was $1~\si{GHz}$ with a main increment $20~\si{MHz}$~\cite{liu2018detectionSubsurfaceFDAMIMO}, and the six main frequencies were $1.00$--$1.10~\si{GHz}$. The controlled increment scan was $\{0,5,10,20,40\}~\si{MHz}$. These settings provide an engineering-anchored, controlled frequency scale for the present study and are not asserted to cover all GPR operating regimes. The common settings are listed in \cref{tab:common_experimental_configuration}.

\begin{table}[!t]
	\centering
	\caption{Common analytical and full-wave configuration.}
	\label{tab:common_experimental_configuration}
	\begin{tblr}{
			width=\columnwidth,
			colspec={X[1.4] X[1.8]},
			row{1}={font=\bfseries},
			cells={font=\footnotesize}
		}
		\toprule
		Configuration item                      & Value                                                                                      \\
		\midrule
		Events and receivers                    & $N_e=6$, $N_r=8$, $N_eN_r=48$                                                              \\
		Stacking and spacing                    & receiver-fastest/event-slowest; $\SI{0.10}{m}$                                             \\
		Reference, main increment, and main set & $f_0=\SI{1}{GHz}$; $\Delta f=\SI{20}{MHz}$; six frequencies from $1.00$ to $1.10~\si{GHz}$ \\
		Increment scan                          & $\{0,5,10,20,40\}~\si{MHz}$                                                                \\
		Media, correlation, and uncertainty     & S1--S5; base spatial correlation $\SI{0.30}{m}$; multipliers $\{0.5,1,2\}$                 \\
		Random-field sampler                     & Gaussian latent field; $\ell_s=\SI{0.30}{m}$; Cholesky jitter $10^{-10}$; seed $2026071001$ \\
		Training, validation, and test          & $256/128/128$; analytical ensemble $=512$                                                  \\
		\bottomrule
	\end{tblr}
\end{table}

The five physical backgrounds were dry lunar soil (S1), basalt (S2), pure water ice (S3), water-bearing kaolinite (S4), and fine-grained clay (S5). Their nominal states and literature sources are listed in \cref{tab:medium_catalog}. Each state was represented by a five-parameter Cole--Cole vector in $\mathcal M_{\mathrm{CC}}$. For reproducibility, the fields were sampled in transformed coordinates $\bm z=\mathcal T(\bm\mu)$, with
\begin{equation}
\begin{aligned}
\mathcal T(\bm\mu)&=\left[\log\varepsilon_\infty,\log\max(\Delta\varepsilon,\sigma_f),\log\tau,\right.\\
&\qquad\left.\log\frac{\alpha}{1-\alpha},\log(\sigma+\sigma_f)\right]^{\mathrm T},
\qquad \sigma_f=10^{-14},
\end{aligned}
\label{eq:latent_cole_cole_transform}
\end{equation}
where boundary values of $\alpha$ were clipped to the floating-point interior. On the $12\times12$ spatial grid, the Gaussian covariance and its regularized Cholesky factor were
\begin{equation}
\begin{aligned}
C_{s,ij}&=\exp\left(-\frac{\lVert\bm x_i-\bm x_j\rVert_2^2}{2\ell_s^2}\right),\\
L_s&=\operatorname{chol}(\mathbf C_s+10^{-10}\mathbf I),
\qquad \ell_s=\SI{0.30}{m}.
\end{aligned}
\label{eq:latent_spatial_covariance}
\end{equation}
The physical component scales were
\begin{equation}
\begin{aligned}
\bm s_\mu&=u\,\bm b\odot(1+0.2\bm q),\\
&\qquad u\in\{0.5,1,2\},\\
\bm b&=(b_1,b_2,b_3,b_4,b_5),\\
b_1&=\max(0.05\varepsilon_\infty,0.01),\\
b_2&=\max(0.05\Delta\varepsilon,0.002),\\
b_3&=\max(0.10\tau,10^{-14}),\\
b_4&=0.25\min(\alpha,1-\alpha),\\
b_5&=\max(0.10\sigma,0),\\
\bm d&=(1,1,10^{-12},1,10^{-12}),\\
\bm q&=\operatorname{clip}\left(\frac{|\bm\mu_b-\bm\mu_r|}{\bm d},0,5\right).
\end{aligned}
\label{eq:latent_parameter_scale}
\end{equation}
Each sample used independent standard-normal innovations scaled by the corresponding latent-coordinate deviations and mapped back to physical coordinates. The inverse transform enforced the admissible Cole--Cole bounds. A deterministic seed was derived from the master seed, scene ID, sample ID, and stream ID through SHA-256 before initializing Mersenne Twister. The 512 samples used IDs $1$--$256$, $257$--$384$, and $385$--$512$ for training, validation, and test, respectively. Training samples were used for moments and risk, validation samples for frozen selection, and test samples for independent evaluation.

\begin{table*}[!t]
	\centering
	\caption{Source-cited Cole--Cole parameters of the physical-background catalogue. The listed $\varepsilon_s$ values satisfy $\varepsilon_s=\varepsilon_\infty+\Delta\varepsilon$ and are included for reproducibility.}
	\label{tab:medium_catalog}
	\begin{tblr}{
			width=\textwidth,
			colspec={X[0.45] X[1.8] X[0.65] X[0.8] X[0.85] X[1.0] X[0.55] X[0.9]},
			row{1}={font=\bfseries},
			cells={font=\scriptsize}
		}
		\toprule
		State & Medium type                                                                                    & $\varepsilon_s$ & $\varepsilon_\infty$ & $\Delta\varepsilon$ & $\tau$ (s)           & $\alpha$ & $\sigma$ (S/m)      \\
		\midrule
		S1    & Dry lunar soil \cite{strangway1974electricalLunarSoil,olhoeft1974electricalPropertiesApollo15} & 3.05            & 3.00                 & 0.05                & $1.0\times10^{-6}$   & 0.30     & $1.0\times10^{-14}$ \\
		S2    & Basalt \cite{olhoeft1973electricalPropertiesLunarBasalt}                                       & 1000            & 8.0                  & 992.0               & $1.0\times10^{-6}$   & 0.30     & $1.0\times10^{-8}$  \\
		S3    & Pure water ice \cite{auty1952dielectricPropertiesIce}                                          & 91.0            & 3.15                 & 87.85               & $2.5\times10^{-5}$   & 0        & $1.0\times10^{-8}$  \\
		S4    & Water-bearing kaolinite \cite{mansour2020dielectricKaoliniteWaterSediment}                     & 35.6            & 2.0                  & 33.6                & $5.0\times10^{-12}$  & 0.20     & $8.0\times10^{-2}$  \\
		S5    & Fine-grained clay \cite{schwing2013dielectricClaySoil}                                         & 30.26           & 10.7                 & 19.56               & $9.55\times10^{-12}$ & 0.062    & 0                   \\
		\bottomrule
	\end{tblr}
\end{table*}

The controlled reference families are summarized in \cref{tab:reference_ensemble}; $\bm z=\mathcal T(\bm\mu)$ denotes the latent coordinate, $\bm z_e$ the same-material endpoint, and $\mathcal S$ the closed S1--S5 catalogue. The table records construction and mismatch conditions without changing the sampled ensemble.

\begin{table*}[!t]
	\centering
	\caption{Controlled reference-state families, latent constructions, and mismatch conditions.}
	\label{tab:reference_ensemble}
	\begin{tblr}{
			width=\textwidth,
			colspec={X[0.8] X[1.1] X[2.1]},
			row{1}={font=\bfseries},
			cells={font=\scriptsize}
		}
		\toprule
		Reference family & Construction                         & Mismatch coordinate and theoretical construction \\
		\midrule
		R0               & Matched reference                      & $\rho=0$; $\bm z_r=\bm z_b$ \\
		R1/R2            & Same-material latent path             & $\begin{gathered}\bm z_r(\rho)=\bm z_b+\rho(\bm z_e-\bm z_b),\\ \rho=1,2\end{gathered}$ \\
		R3               & Same material with spatial smoothing  & $\begin{gathered}w_{ij}(\ell_R)\propto\exp[-\lVert\bm x_i-\bm x_j\rVert_2^2/(2\ell_R^2)],\ \sum_jw_{ij}=1,\\ \bm z_{r,i}=\sum_jw_{ij}\bm z_{b,j}+\bm z_r-\bm z_b,\quad \ell_R\in\{0.3,0.6,0.9\}~\si{m}\end{gathered}$ \\
		R4               & Nearest wrong material in S1--S5    & $r^\star=\displaystyle\arg\min_{s\in\mathcal S\setminus\{b\}}\lVert\mathcal T(\bm\mu_s)-\mathcal T(\bm\mu_b)\rVert_2$ \\
		\bottomrule
	\end{tblr}
\end{table*}

For FDTD checks, the source was an ideal $x$-directed Hertzian dipole, the recorded field was point $E_x$, and the polarization was $x/E_x$. A $\SI{1.1}{GHz}$ Ricker pulse, an eight-pole multi-Debye representation, a fifteen-cell perfectly matched layer (PML), and exact-frequency discrete-time Fourier transform (DTFT) extraction were used. These choices retain controlled event semantics across analytical and full-wave calculations. The numerical settings are listed in \cref{tab:common_fdtd_configuration}.

\begin{table}[htbp]
	\centering
	\caption{Common full-wave configuration.}
	\label{tab:common_fdtd_configuration}
	\begin{tblr}{
			width=\columnwidth,
			colspec={X[1.0] X[2.0]},
			row{1}={font=\bfseries},
			cells={font=\footnotesize}
		}
		\toprule
		Item                    & Configuration                                                \\
		\midrule
		Source and field        & ideal $x$-Hertzian dipole; point $E_x$; co-polarized $x/E_x$ \\
		Pulse and dispersion    & $\SI{1.1}{GHz}$ Ricker; eight-pole multi-Debye               \\
		Boundary and extraction & fifteen-cell PML; exact-frequency DTFT                       \\
		Analysis                & six frequencies, identical event semantics and stacking      \\
		\bottomrule
	\end{tblr}
\end{table}

\subsection{Operational validity of the fixed-reference first-order model}\label{subsec:exp_first_order}

For the anomaly-free background, the observation-domain discrepancy was defined as
\begin{equation}
	D_{\mathrm{obs}}=10\log_{10}\frac{\lVert\bm r^{(\mathrm{sh})}\rVert_2^2}{\lVert\bm c^{(\mathrm{sh},\mathrm{ex})}\rVert_2^2},
	\label{eq:experimental_observation_discrepancy}
\end{equation}
and the task-domain discrepancy as
\begin{equation}
	D_A=10\log_{10}\frac{\lVert\mathbf A\bm r^{(\mathrm{sh})}\rVert_2^2}{\lVert\mathbf A\bm c^{(\mathrm{sh},\mathrm{ex})}\rVert_2^2}.
	\label{eq:experimental_task_discrepancy}
\end{equation}
The $0$ dB point denotes equal energy in the numerator and denominator. It is a comparison reference, not a universal DBA threshold. Because the task operator can change the relative projection, both domains were examined.

\begin{figure*}[!t]
	\centering
	\includegraphics[width=0.31\textwidth]{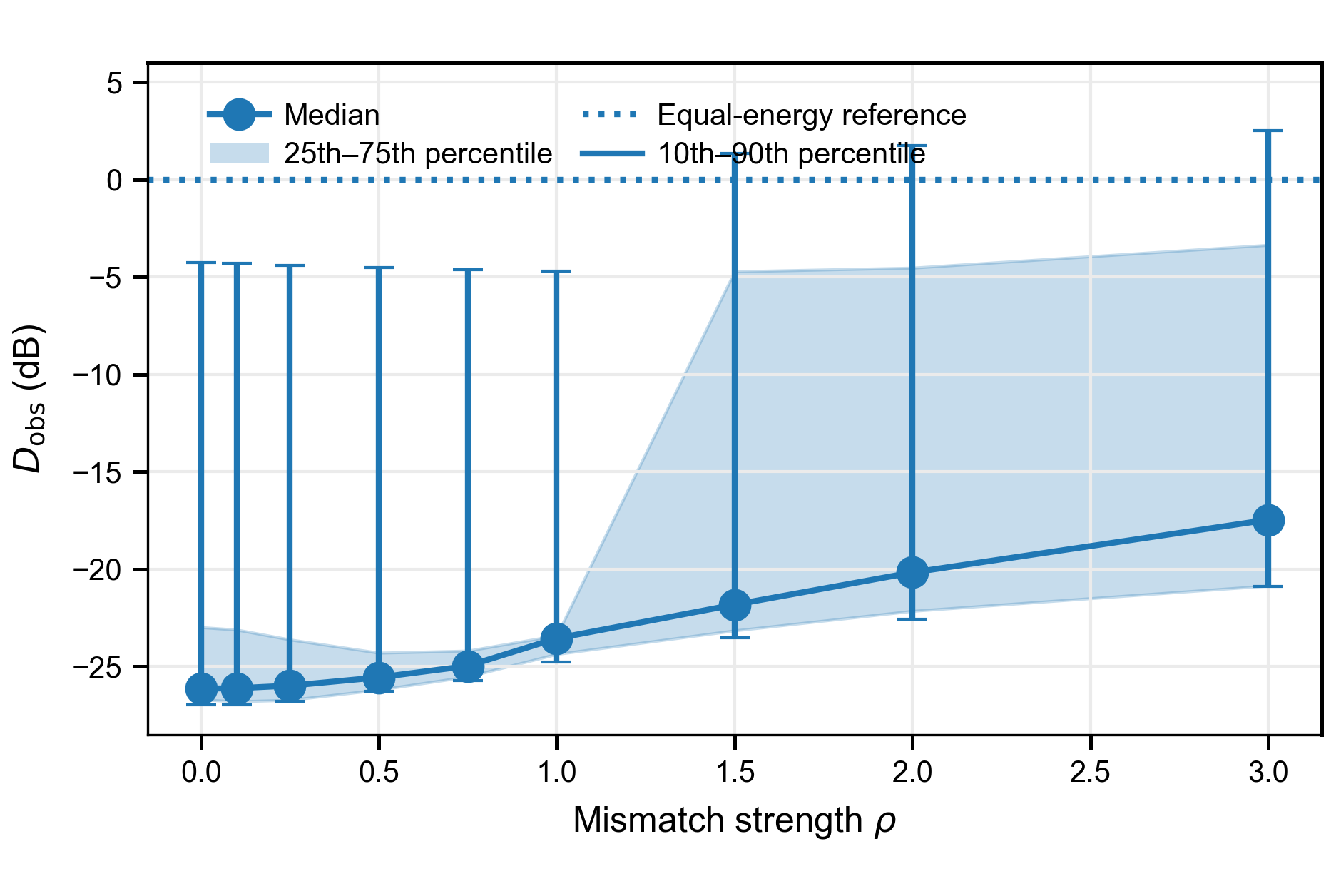}
	\includegraphics[width=0.31\textwidth]{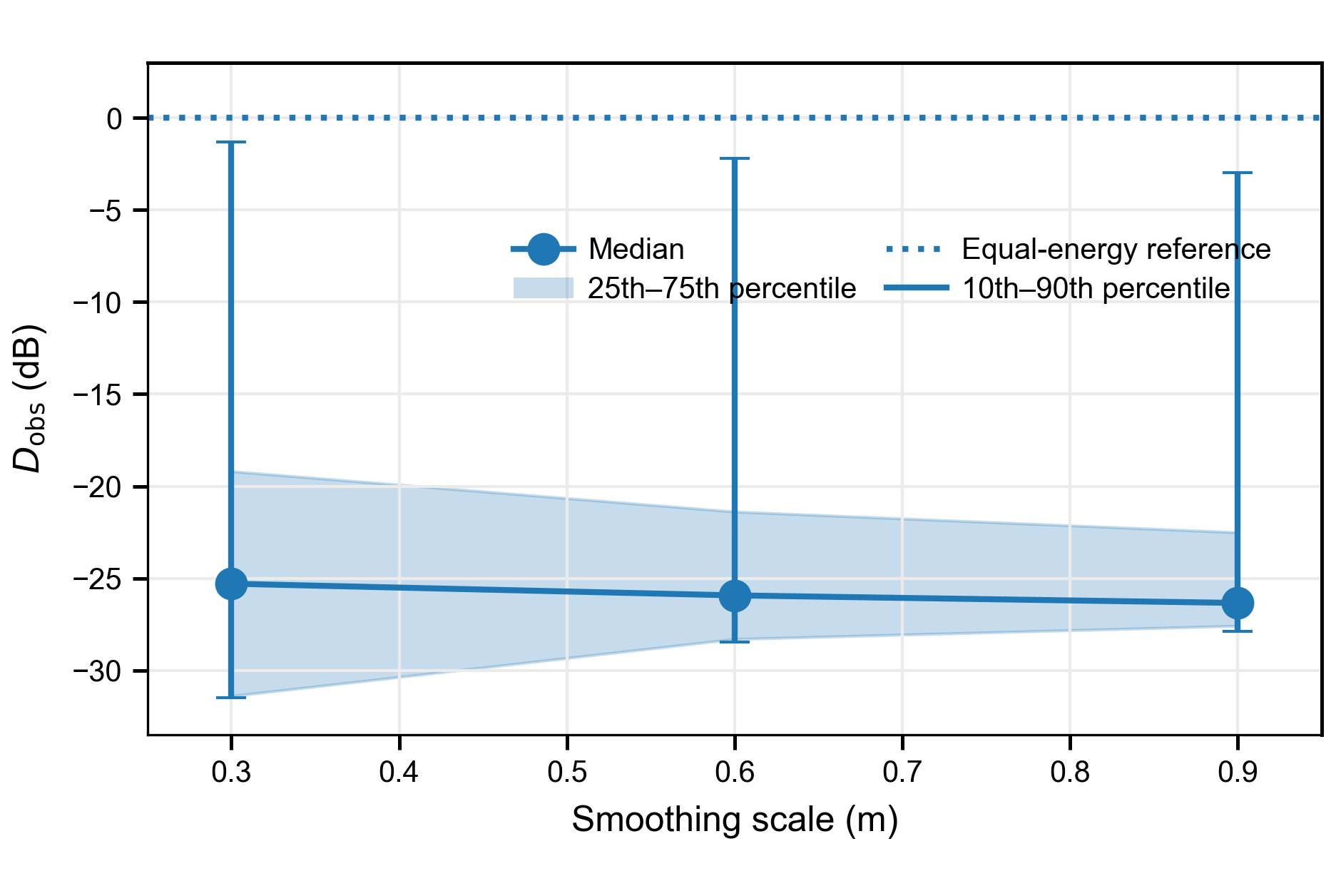}
	\includegraphics[width=0.31\textwidth]{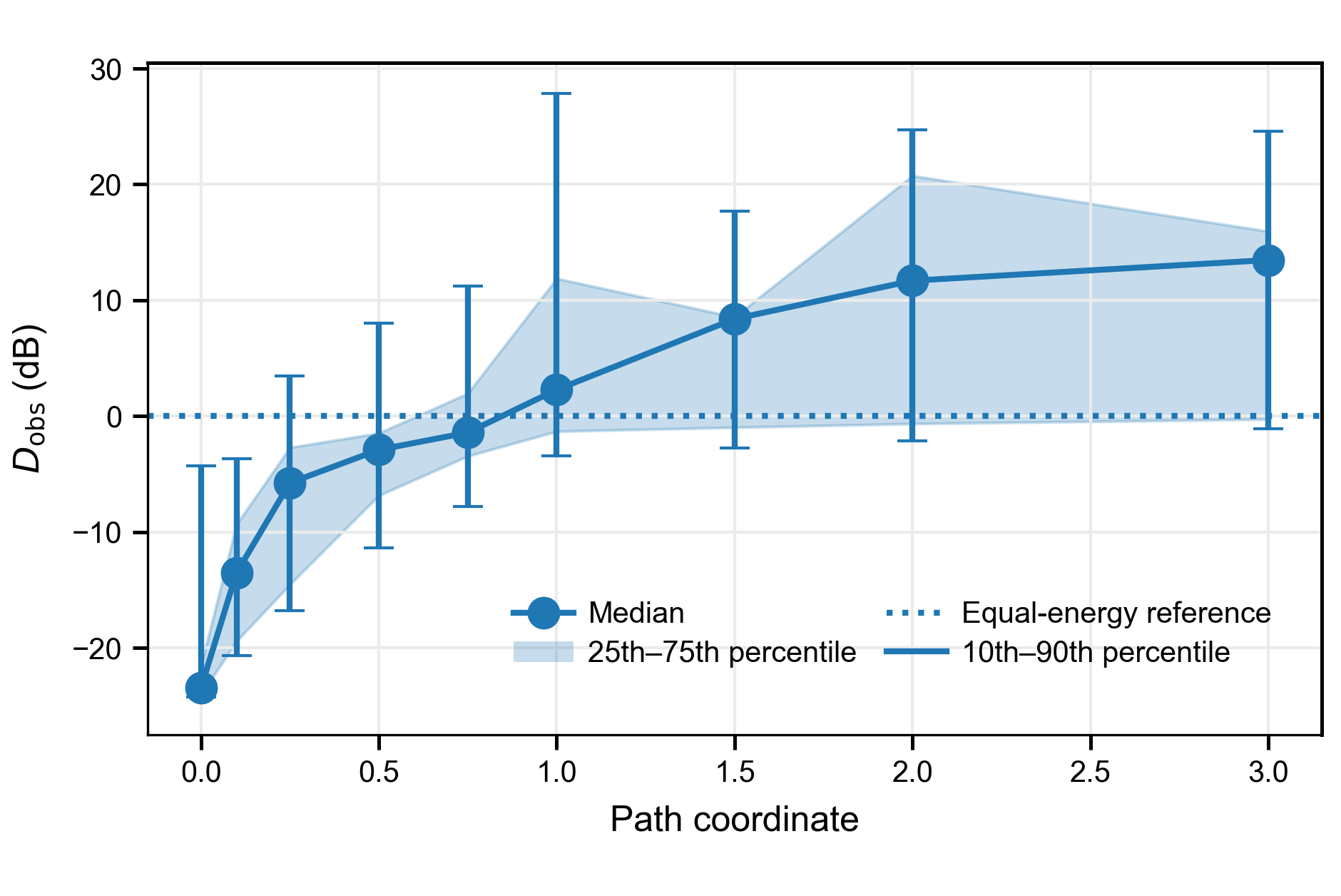}
	\caption{Analytical observation-domain discrepancy envelopes for parameter mismatch, spatial smoothing, and wrong-material references.}
	\label{fig:analytical_observation_envelopes}
\end{figure*}

The analytical sweep used same-material paths $\rho=\{0,0.1,0.25,0.5,0.75,1,1.5,2,3\}$, spatial smoothing scales $0.3$, $0.6$, and $0.9~\si{m}$, and nine coordinates for the wrong-material path. Five media, five frequency increments, and two interfaces produced fifty paired conditions. For the same-material path, the median $D_{\mathrm{obs}}$ was $-25.88~\si{dB}$ at $\rho=0$, $-23.60~\si{dB}$ at $\rho=1$, $-20.21~\si{dB}$ at $\rho=2$, and $-17.50~\si{dB}$ at $\rho=3$. The smoothing medians were $-25.71$, $-26.15$, and $-26.52~\si{dB}$ for $0.3$, $0.6$, and $0.9~\si{m}$. The wrong-material path degraded more rapidly, with medians of $-13.58~\si{dB}$ at coordinate $0.1$, $-5.78~\si{dB}$ at $0.25$, $-1.42~\si{dB}$ at $0.75$, and $2.28~\si{dB}$ at $1$. The wrong-material coordinate is not equivalent to the same-material parameter $\rho$. The resulting envelopes are shown in \cref{fig:analytical_observation_envelopes}.

\begin{figure}[htbp]
	\centering
	\includegraphics[width=\columnwidth]{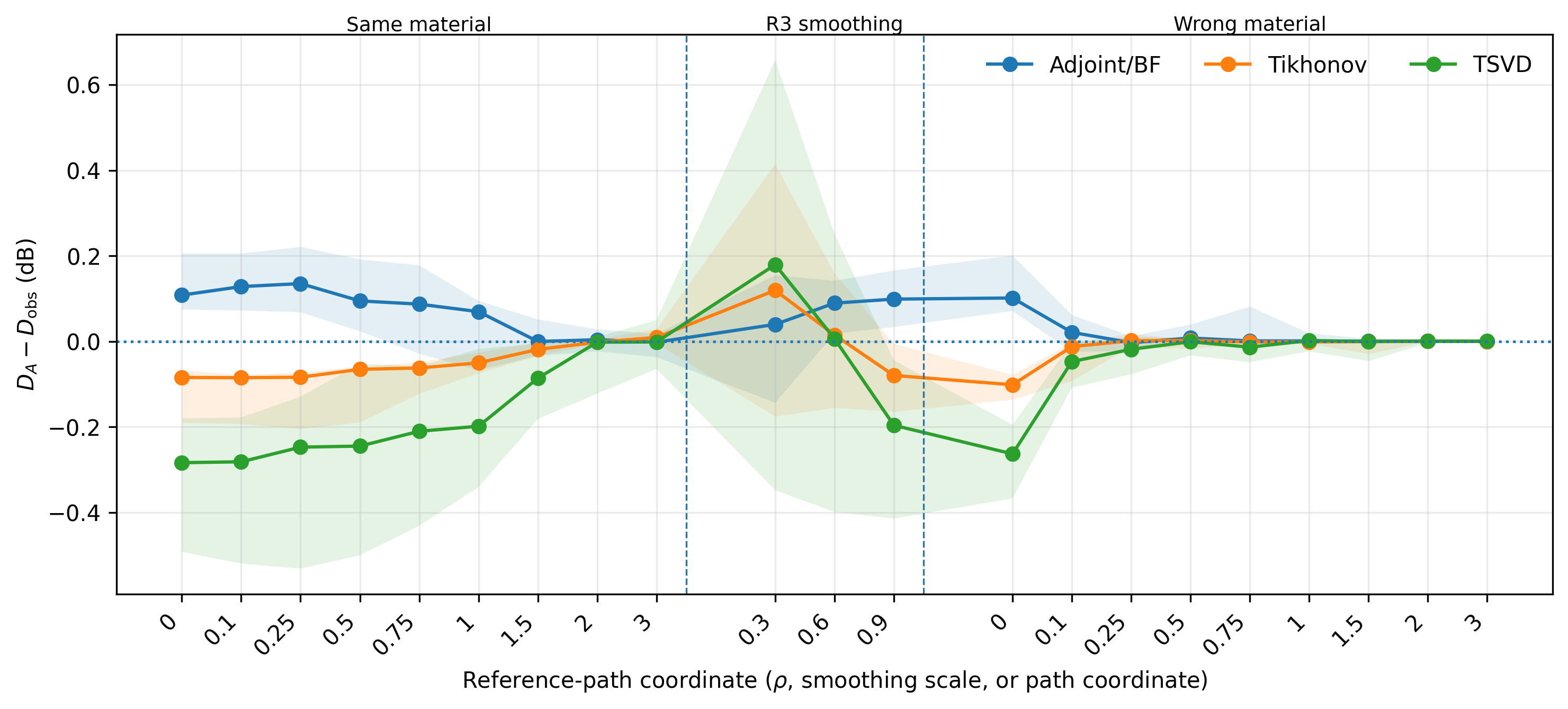}
	\caption{Task-domain changes relative to the observation-domain discrepancy for the three frozen task operators.}
	\label{fig:analytical_task_projection_shift}
\end{figure}

Across the three frozen task operators, the maximum median change in $D_A-D_{\mathrm{obs}}$ was $0.08$, $0.18$, and $0.01~\si{dB}$ for Adjoint/BF, Tikhonov, and TSVD, respectively. The maximum increase over all conditions was $0.83~\si{dB}$, and the Spearman association with the observation-domain ordering exceeded $0.99$. This agreement is restricted to the three operators and the tested conditions. These task projection shifts provide the experimental check of \cref{prop:task_discrepancy_transfer} for the analytical sweep. The task projection shifts are shown in \cref{fig:analytical_task_projection_shift}.

\begin{figure}[!t]
	\centering
	\includegraphics[width=0.48\columnwidth]{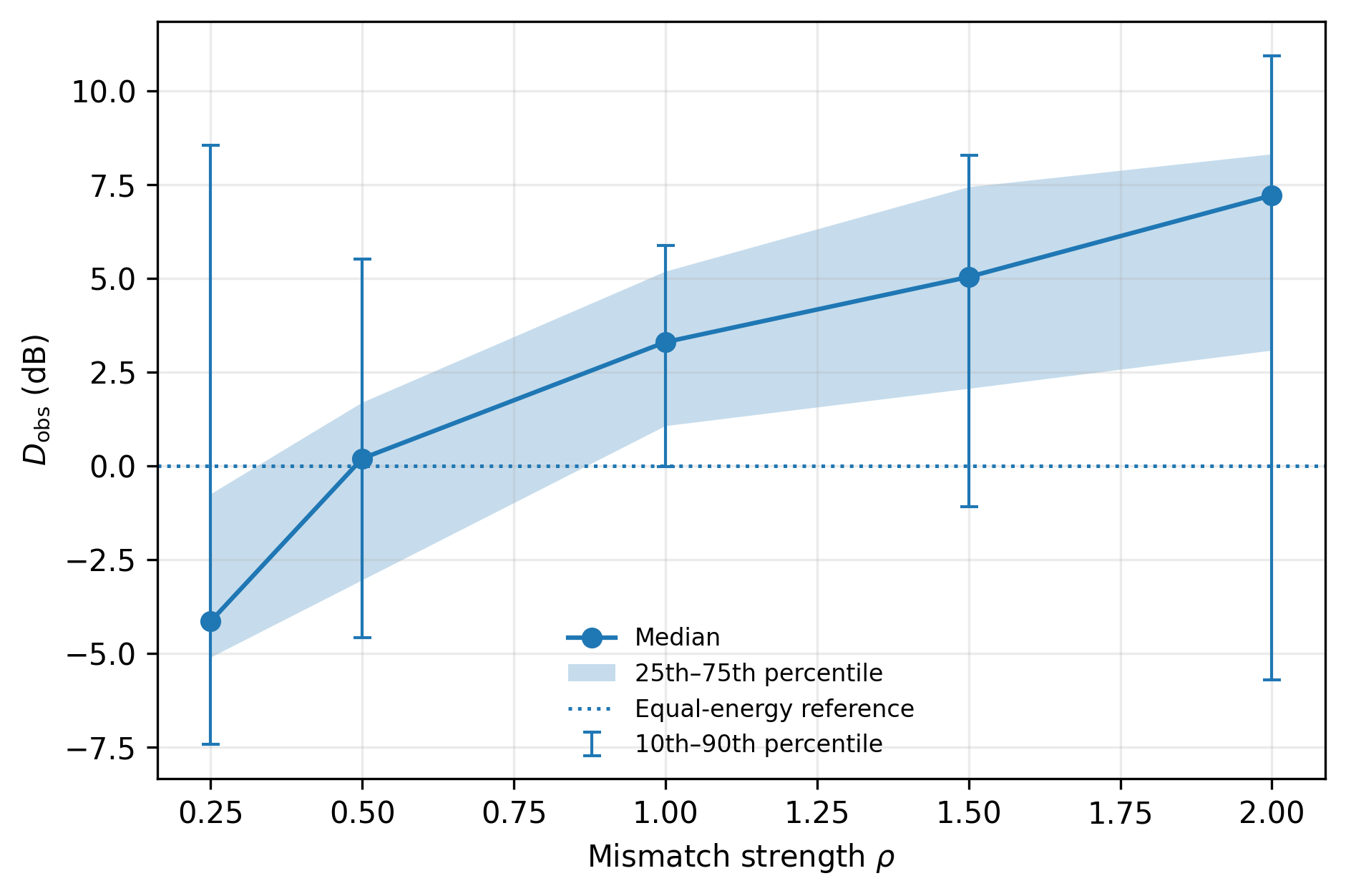}
	\includegraphics[width=0.48\columnwidth]{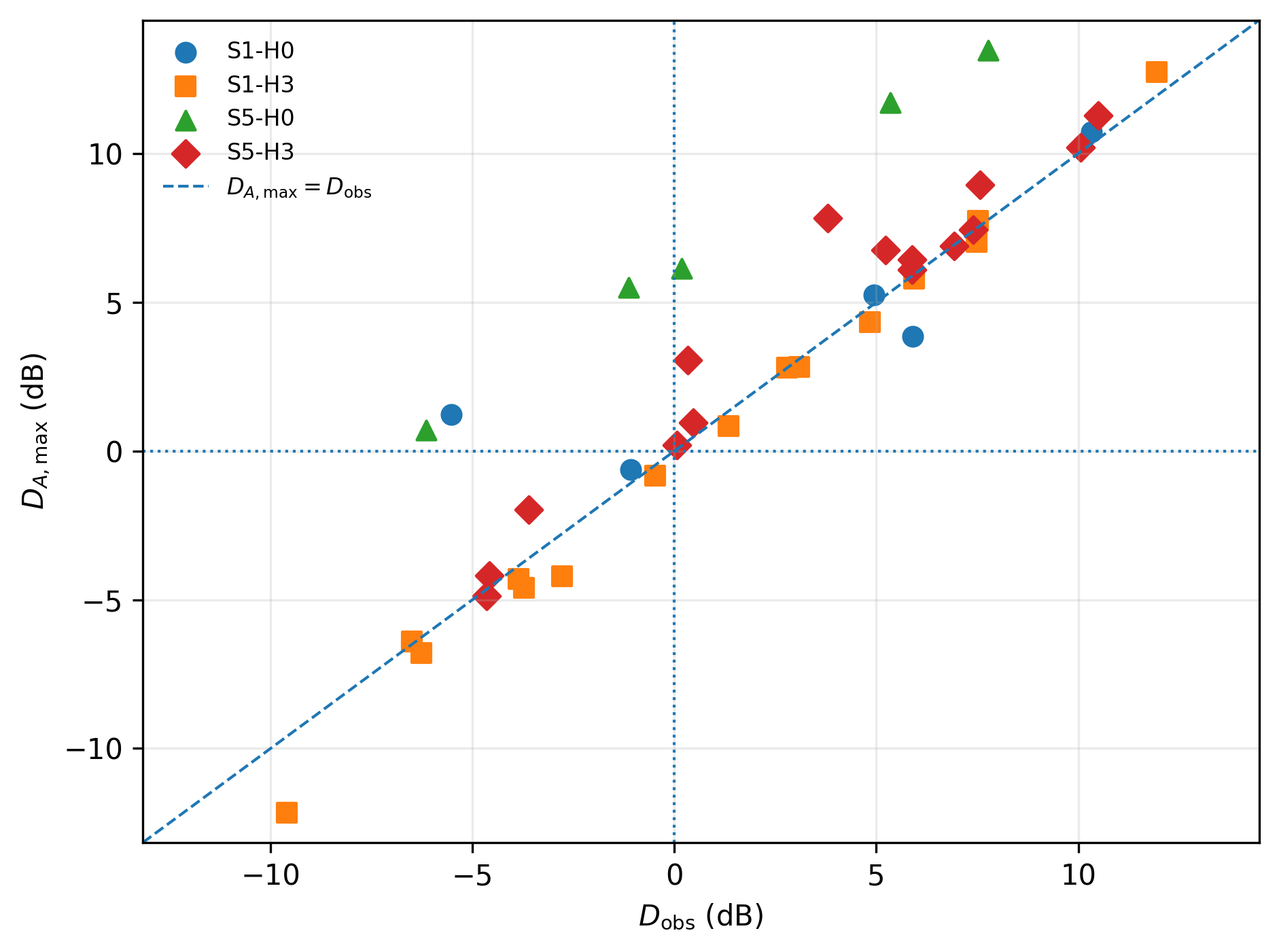}
	\caption{Full-wave checks of observation-domain discrepancy and its task-domain transfer.}
	\label{fig:fullwave_first_order_stress}
\end{figure}

The full-wave check covered S1 and S5, smooth and combined interfaces H0 and H3, and $\rho=0.25$, $0.5$, $1$, $1.5$, and $2$, giving forty records. A finite-difference tangent with $h=0.025$ defined the first-order comparison. The median $D_{\mathrm{obs}}$ increased from $-4.15~\si{dB}$ at $\rho=0.25$ to $0.20$, $3.30$, $5.04$, and $7.21~\si{dB}$ at the remaining mismatch levels. In twenty-nine of forty records, the maximum task-domain remainder reached or exceeded the full task response. The median of $D_{A,\max}-D_{\mathrm{obs}}$ was $0.21~\si{dB}$ with interquartile range $[-0.28,1.40]~\si{dB}$. The full-wave discrepancy was generally larger than the analytical discrepancy, although its direction agreed with the analytical trend. The result limits quantitative first-order interpretation by both reference structure and mismatch scale. The paired checks are shown in \cref{fig:fullwave_first_order_stress}.

\subsection{Joint observation covariance of the shared residual}\label{subsec:exp_observation_covariance}

\begin{figure*}[!t]
	\centering
	\includegraphics[width=0.24\textwidth]{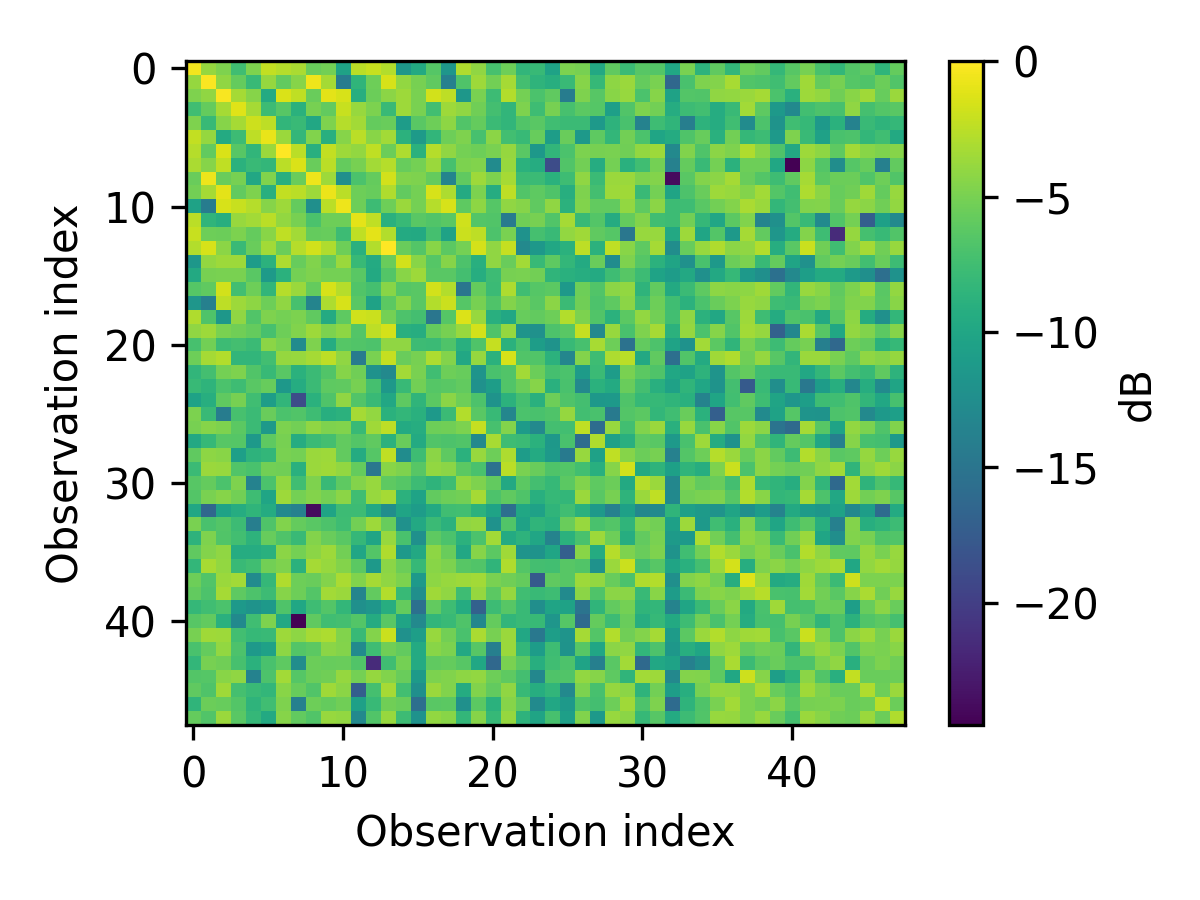}
	\includegraphics[width=0.24\textwidth]{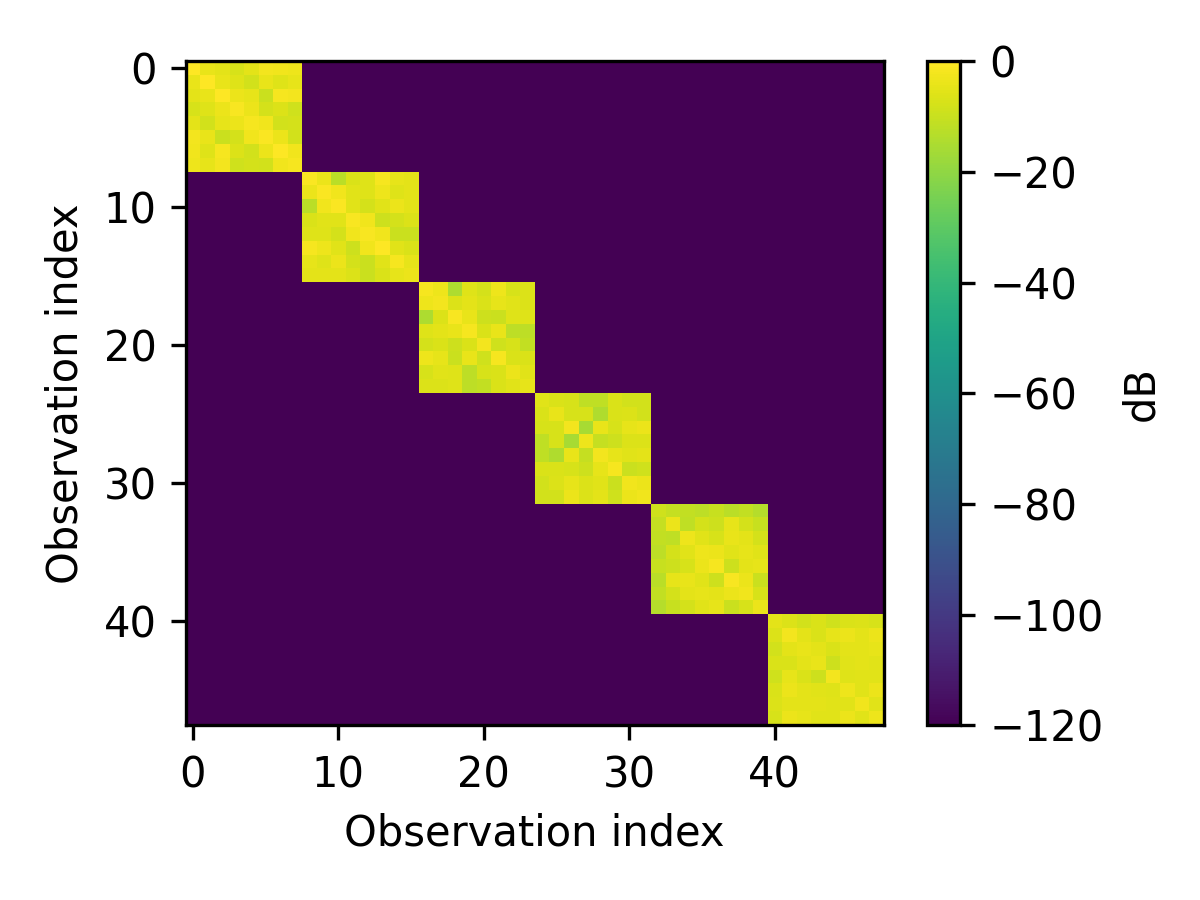}
	\includegraphics[width=0.24\textwidth]{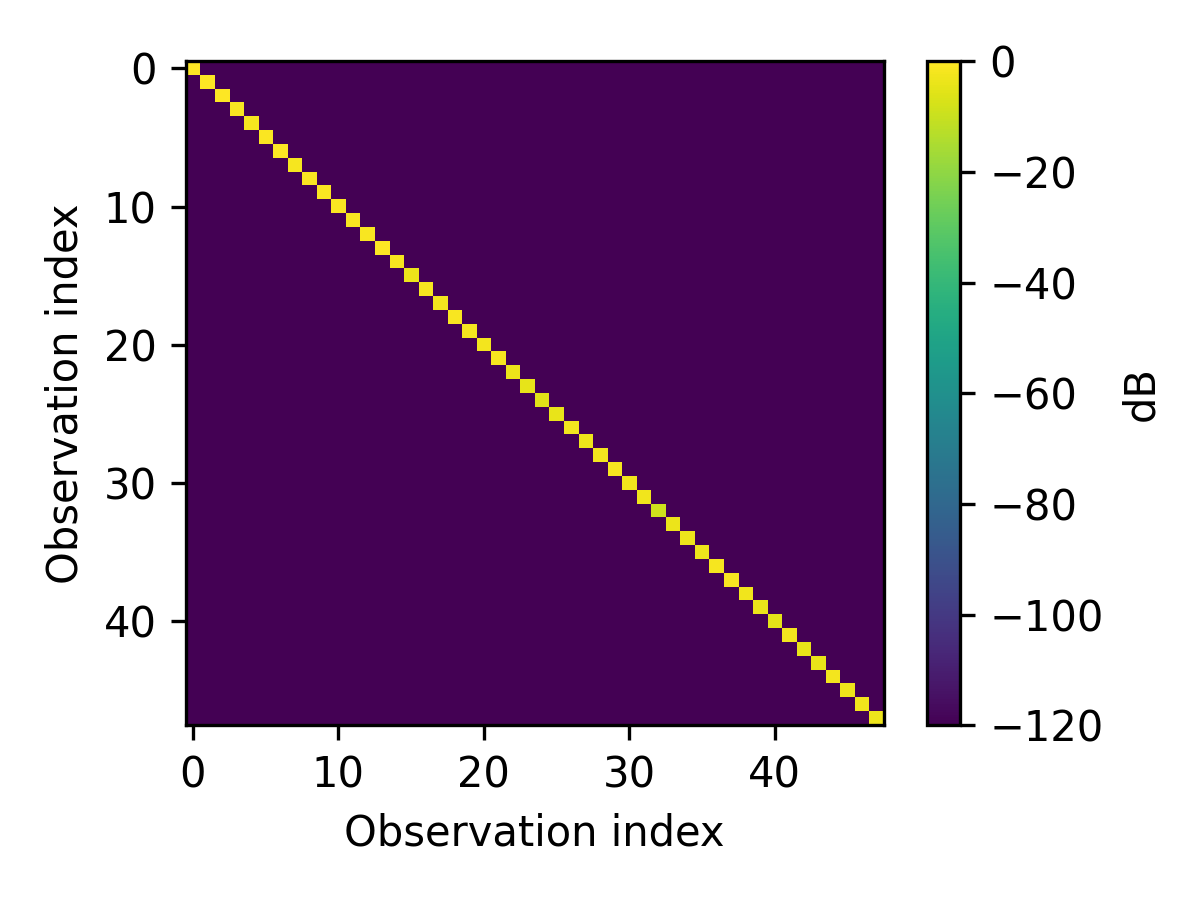}
	\includegraphics[width=0.24\textwidth]{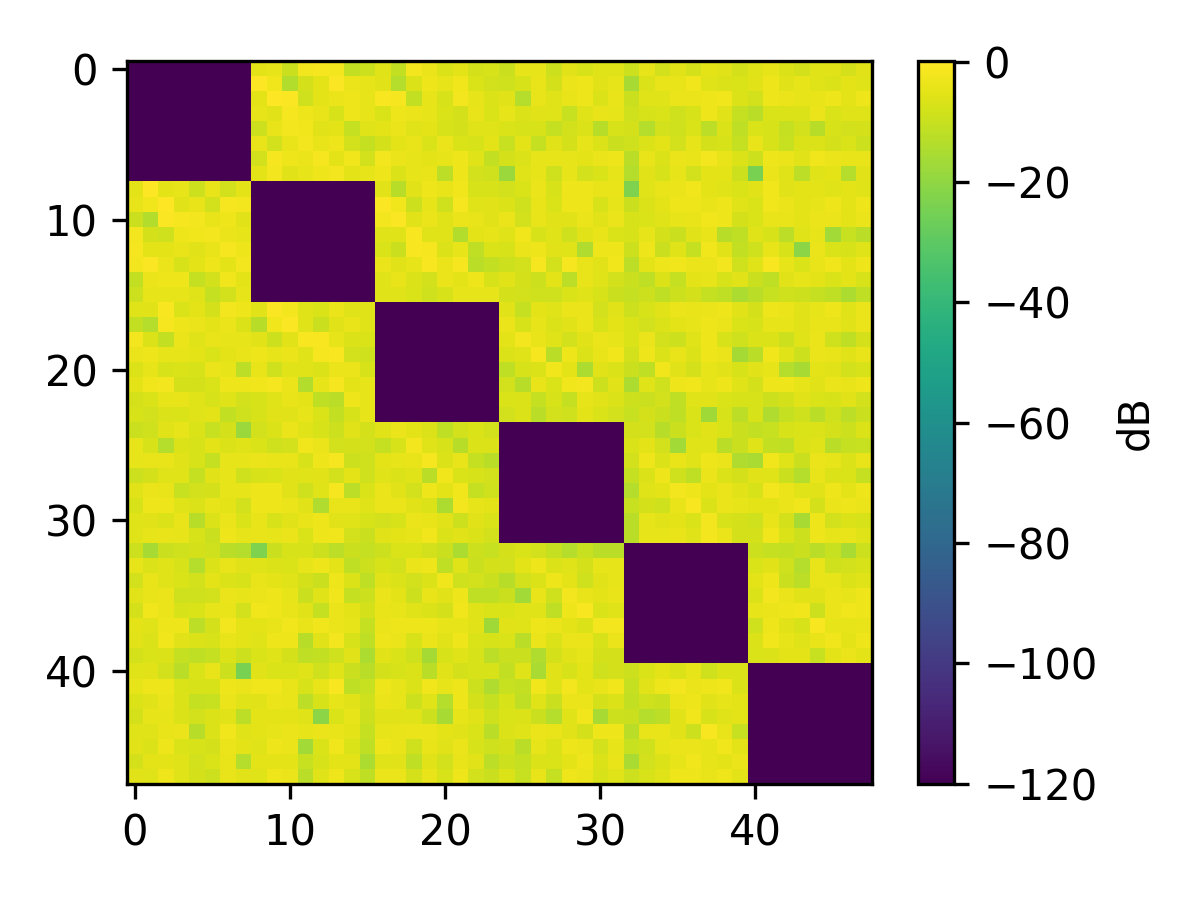}
	\caption{Full, event-block, diagonal, and difference views of the joint residual covariance.}
	\label{fig:joint_covariance_structure}
\end{figure*}

The analytical covariance was computed from $512$ shared residual realizations using the full, event-block, and diagonal models in \eqref{eq:full_covariance_model}--\eqref{eq:diagonal_covariance_model}. The cross-event covariance fractions for S1 and S5 were $0.819$ and $0.866$. Their effective ranks were $2.284$ and $1.708$, with two eigen-directions required to reach $p=0.9$. The full covariance therefore contained substantial off-diagonal event structure under the present residual ensemble. The matrix views are given in \cref{fig:joint_covariance_structure}, and the numerical decomposition is listed in \cref{tab:covariance_decomposition}.

\begin{table}[htbp]
	\centering
	\caption{Analytical covariance decomposition.}
	\label{tab:covariance_decomposition}
	\footnotesize
	\begin{tblr}{
			width=\columnwidth,
			colspec={X[0.8] X[1.4] X[1.1] X[0.8]},
			row{1}={font=\bfseries},
			cells={font=\footnotesize}
		}
		\toprule
		Medium & Cross-event fraction & Effective rank & $p_{0.9}$ \\
		\midrule
		S1     & 0.819                & 2.284          & 2         \\
		S5     & 0.866                & 1.708          & 2         \\
		\bottomrule
	\end{tblr}
\end{table}

\begin{figure}[htbp]
	\centering
	\includegraphics[width=\columnwidth]{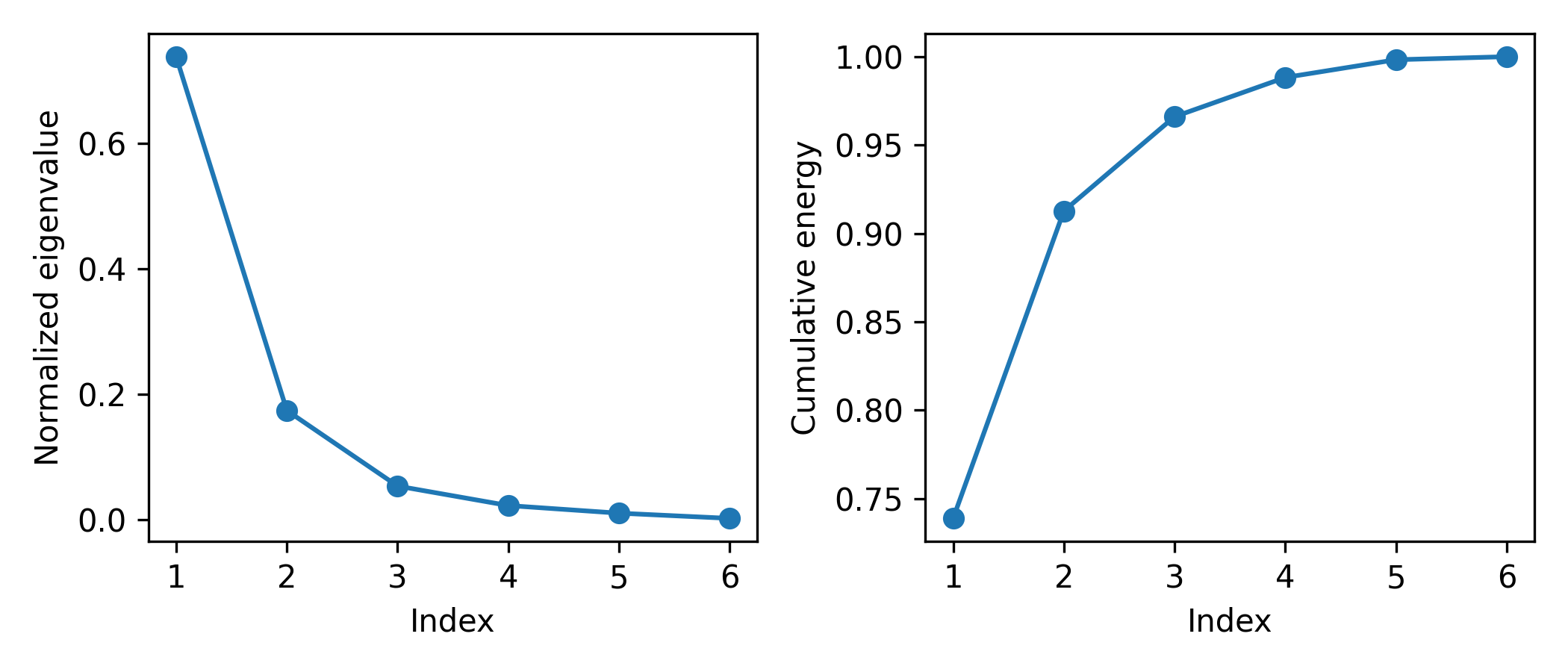}
	\caption{Covariance spectra for representative physical backgrounds.}
	\label{fig:covariance_spectrum}
\end{figure}

The covariance spectrum of S1 placed $73.88\%$ of the energy in the first eigenvalue and $91.27\%$ in the first two. S5 placed $83.12\%$ in the first eigenvalue. These values describe the tested ensemble and do not equate effective rank with algebraic rank. The spectra are shown in \cref{fig:covariance_spectrum}.

The compact full-wave check used four scene--propagation combinations and three realizations. At zero and $20~\si{MHz}$ increments, the paired cross-event fractions were $0.845\mathbin{\to}0.816$, $0.884\mathbin{\to}0.865$, $0.900\mathbin{\to}0.813$, and $0.831\mathbin{\to}0.827$. The range $0.813$--$0.900$ supports the existence of joint structure under more complete propagation, but the small number of realizations does not support a population covariance estimate.

\iffalse
	The compact direction check is shown in \cref{fig:compact_fdtd_covariance}.
	\begin{figure}[htbp]
		\centering
		\includegraphics[width=\columnwidth]{fig_compact_fdtd_covariance.png}
		\caption{Compact full-wave direction check of event covariance.}
		\label{fig:compact_fdtd_covariance}
	\end{figure}
\fi

\subsection{FDA frequency states and assignment effects}\label{subsec:exp_fda}

\begin{figure}[htbp]
	\centering
	\includegraphics[width=\columnwidth]{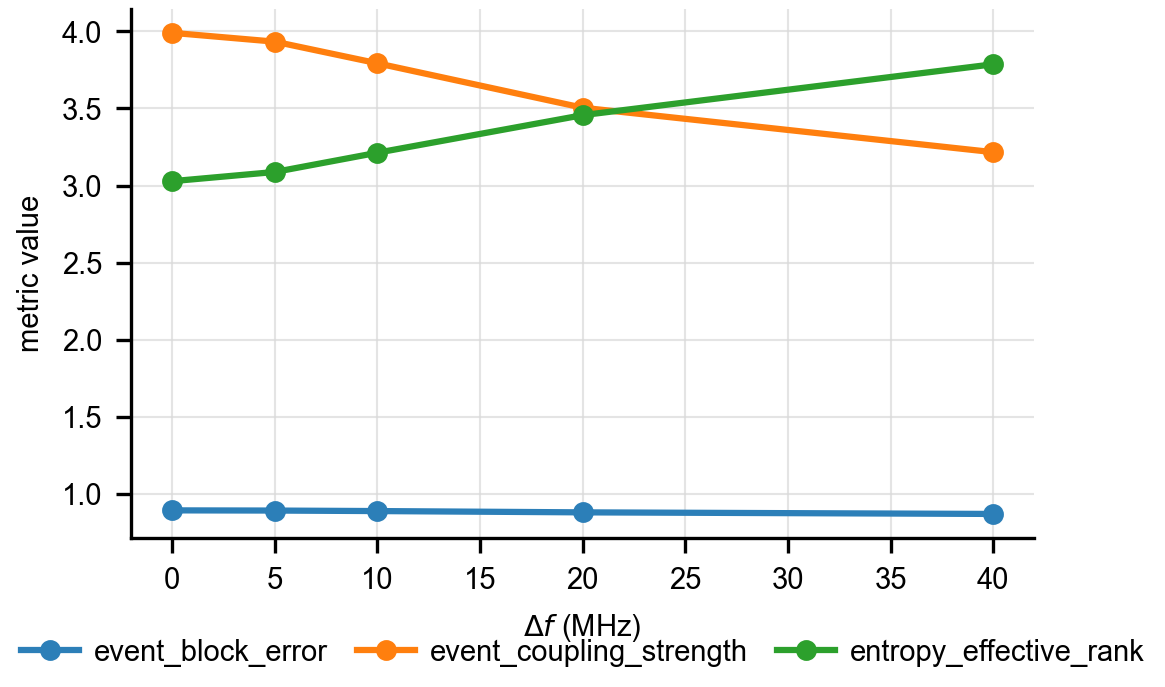}
	\caption{Organization of the shared residual under the controlled frequency-increment scan.}
	\label{fig:fda_frequency_state}
\end{figure}

\begin{figure}[htbp]
	\centering
	\includegraphics[width=\columnwidth]{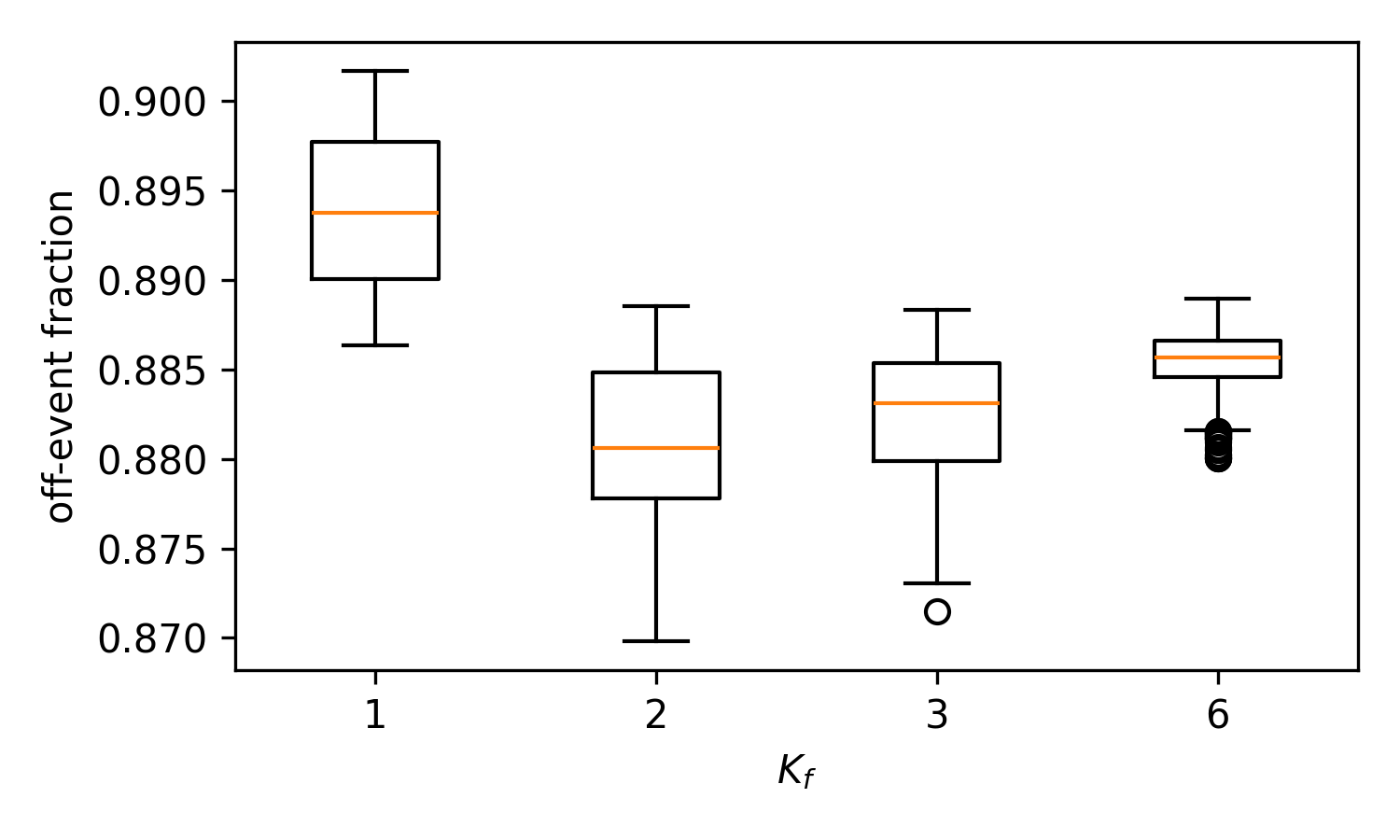}
	\caption{Distribution of assignment-dependent covariance summaries.}
	\label{fig:fda_assignment_distribution}
\end{figure}

The frequency scan used $\Delta f=\{0,5,10,20,40\}~\si{MHz}$ and $K_f=1,2,3,6$ feasible one-hot assignments. The number of assignments was $1$, $20$, $90$, and $720$, respectively. The zero-increment condition retained shared residual structure, confirming that FDA frequency diversity is not the physical source of the covariance. Frequency states changed the spectral organization, event-block pattern, and effective-rank manifestation through $\mathbf L(\mathbf Z)$. The frequency-state and assignment summaries are shown in \cref{fig:fda_frequency_state,fig:fda_assignment_distribution}.

For the $K_f=6$ linear FDA assignment, the cross-event covariance percentiles were $92.361\%$ for S1, $73.611\%$ for S4, and $6.806\%$ for S5. The distributions overlapped across assignments and the ordering changed with the medium condition. The reported percentiles are listed in \cref{tab:fda_assignment_percentile}.

\begin{table}[htbp]
	\centering
	\caption{Cross-event covariance percentiles for the $K_f=6$ linear FDA assignment.}
	\label{tab:fda_assignment_percentile}
	\footnotesize
	\begin{tblr}{
			width=\columnwidth,
			colspec={X[1.2] X[0.9] X[0.9] X[0.9]},
			row{1}={font=\bfseries},
			cells={font=\footnotesize}
		}
		\toprule
		Medium     & S1       & S4       & S5      \\
		\midrule
		Percentile & 92.361\% & 73.611\% & 6.806\% \\
		\bottomrule
	\end{tblr}
\end{table}

The compact full-wave direction check used three realizations for each of four scene--interface combinations. The off-event fractions at $\Delta f=0$ were $0.8447$, $0.8842$, $0.9001$, and $0.8303$ for S1-H0, S1-H3, S5-H0, and S5-H3, respectively. At $\Delta f=20~\si{MHz}$, the corresponding fractions were $0.8161$, $0.8659$, $0.8130$, and $0.8267$. The paired changes were $-0.0286$, $-0.0183$, $-0.0871$, and $-0.0036$. All four conditions retained nonzero off-event structure, while the magnitude of the change depended on the scene and interface. The largest reduction occurred for S5-H0 and the smallest for S5-H3. This finite, low-granularity check supports a directional interpretation in which the frequency increment reorganizes pre-existing shared-residual structure through the propagation mapping.

\subsection{Task transfer, spatial consequences, and reference screening}\label{subsec:exp_task_transfer_screening}

The experimental task component and its risk were defined from the first-order shared response as
\begin{equation}
	\bm e_A^{(\mathrm{sh})}=\bm A\bm c^{(\mathrm{sh})},
	\label{eq:experimental_task_shared_component}
\end{equation}
and
\begin{equation}
	J_A^{(\mathrm{sh})}=\lVert\bm A\bm m_{\mathrm{sh}}\rVert_2^2+
	\operatorname{tr}\left(\bm A\bm R_{\mathrm{sh}}\bm A^{\mathrm H}\right).
	\label{eq:experimental_task_risk_decomposition}
\end{equation}
These definitions calibrate residual energy for the fixed task ports below. They do not constitute a general image-reliability or target-detection metric.
The task-domain discrepancy results reported in \cref{subsec:exp_first_order}, including the $D_A-D_{\mathrm{obs}}$ projection shifts and the full-wave remainder comparison, provide the experimental check of \cref{prop:task_discrepancy_transfer}.

Three fixed task ports were used: direct adjoint projection or beamforming (Adjoint/BF), zero-order Tikhonov, and truncated singular-value decomposition (TSVD). Their regularized inverse conventions follow standard treatments of discrete ill-posed problems \cite{tikhonov1977solution,hansen1992analysisLcurve,hansen1993useLcurve,hansen1998rankDeficient,hansen2010discreteInverse}. The Tikhonov parameter was $\lambda/\sigma_{\max}=30.0$. TSVD retained $r=4$ modes under an energy threshold of $0.4$. The validation residual samples were 257--384, and the maximum numerical closure error was below $2.8\times10^{-16}$. The operator definitions are listed in \cref{tab:fixed_task_operators}.

\begin{table}[htbp]
	\centering
	\caption{Fixed task operators used for task-domain transfer.}
	\label{tab:fixed_task_operators}
	\begin{tblr}{
			width=\columnwidth,
			colspec={X[1.0] X[2.0]},
			row{1}={font=\bfseries},
			cells={font=\footnotesize}
		}
		\toprule
		Task port  & Fixed configuration                                          \\
		\midrule
		Adjoint/BF & Direct adjoint projection                                    \\
		Tikhonov   & Zero-order regularized inverse; $\lambda/\sigma_{\max}=30.0$ \\
		TSVD       & Truncated inverse; $r=4$, energy threshold $0.4$             \\
		\bottomrule
	\end{tblr}
\end{table}

Across sixty independent scene--reference--target test conditions, the risk calibration slopes were $0.983$--$1.010$, Pearson correlations were $0.988$--$0.996$, and Spearman correlations were $0.994$--$0.998$. Mean relative calibration errors for Adjoint/BF, Tikhonov, and TSVD were $0.118$, $0.133$, and $0.137$. These calibration results provide the experimental check of \cref{prop:task_moment_transfer}. The calibration statistics are summarized in \cref{tab:task_risk_calibration_summary}.

\begin{table}[htbp]
	\centering
	\caption{Task-risk calibration summary over independent test conditions.}
	\label{tab:task_risk_calibration_summary}
	\begin{tblr}{
			width=\columnwidth,
			colspec={X[0.8] X[1.2] X[1.0] X[0.7]},
			row{1}={font=\bfseries},
			cells={font=\footnotesize}
		}
		\toprule
		Task       & Slope range  & Pearson range & Spearman range \\
		\midrule
		Adjoint/BF & 0.983--1.010 & 0.988--0.996  & 0.994--0.998   \\
		Tikhonov   & 0.983--1.010 & 0.988--0.996  & 0.994--0.998   \\
		TSVD       & 0.983--1.010 & 0.988--0.996  & 0.994--0.998   \\
		\bottomrule
	\end{tblr}
\end{table}

\begin{figure*}[!t]
	\centering
	\includegraphics[width=0.31\textwidth]{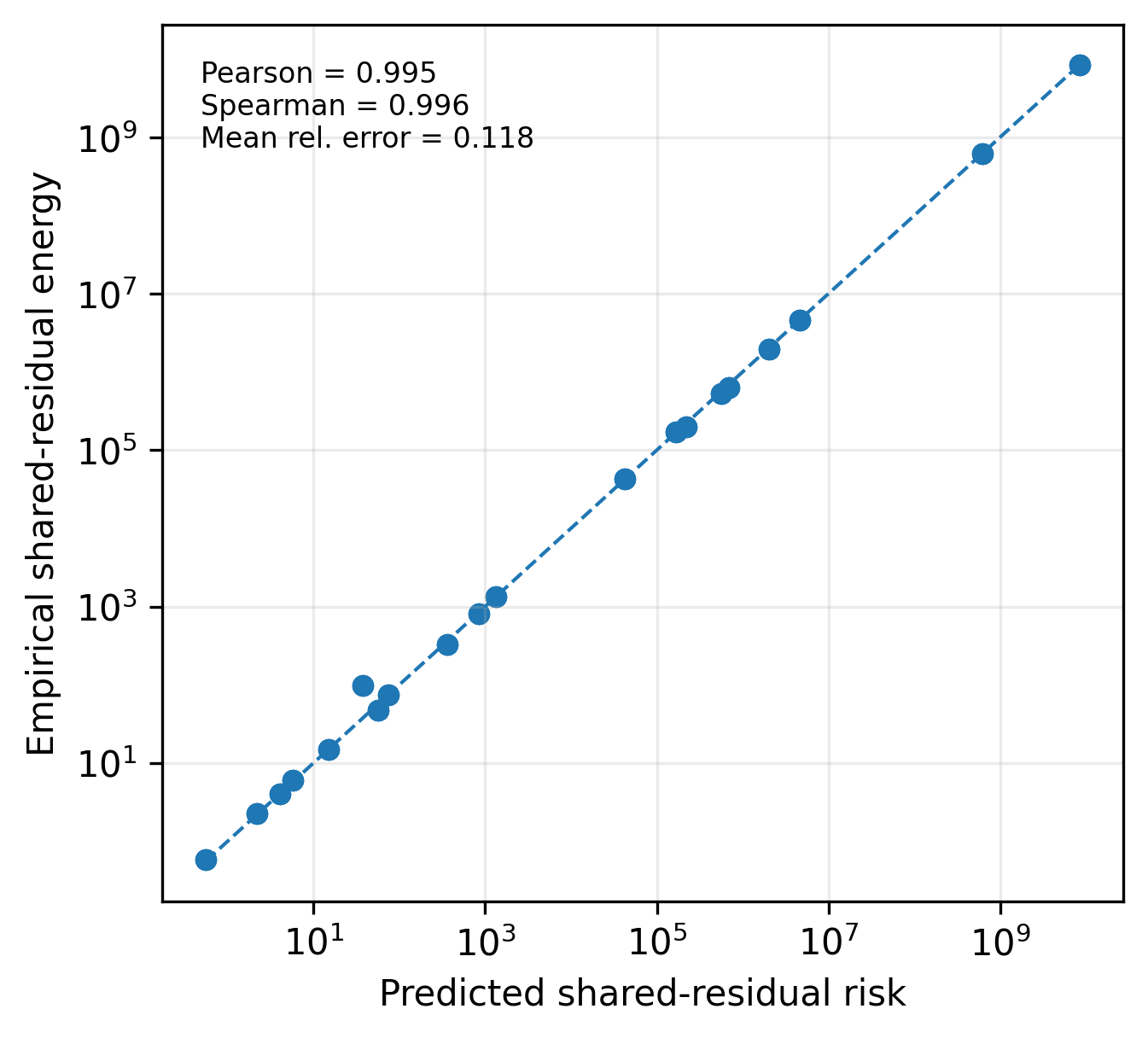}
	\includegraphics[width=0.31\textwidth]{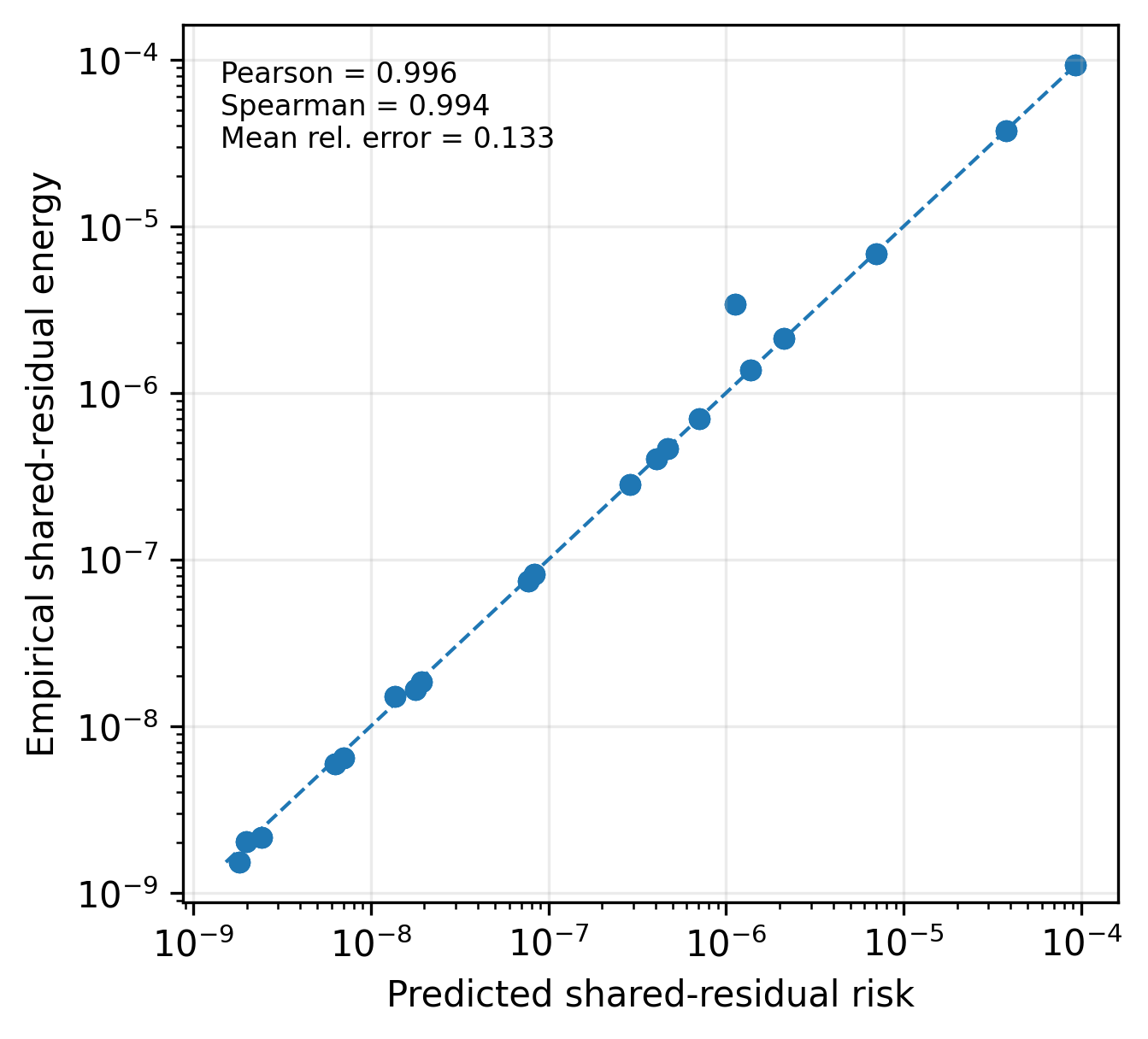}
	\includegraphics[width=0.31\textwidth]{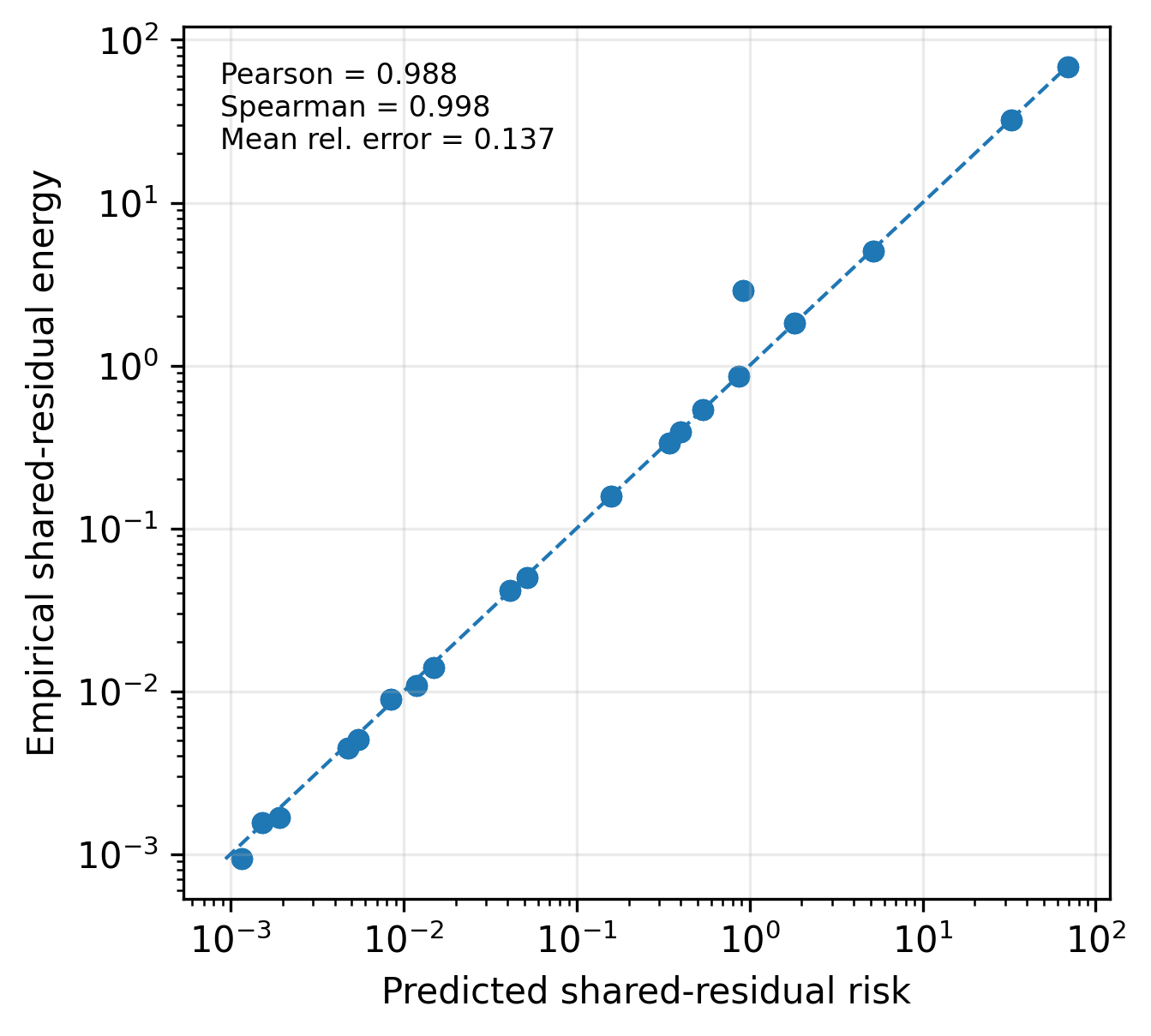}
	\caption{Calibration of mean-plus-covariance task risk for the three fixed task ports.}
	\label{fig:task_risk_calibration}
\end{figure*}

The average mean-risk fractions were $0.5675$, $0.5691$, and $0.5639$ for the three ports, while covariance fractions were $0.4325$, $0.4309$, and $0.4361$. The corresponding median mean fractions were $0.8024$, $0.8058$, and $0.7955$. Both systematic and random terms therefore contributed under the tested conditions. These mean--covariance fractions provide the corresponding experimental decomposition in \cref{prop:task_moment_transfer}. The calibration curves and component fractions are shown in \cref{fig:task_risk_calibration} and listed in \cref{tab:task_risk_component_summary}.

\begin{table}[htbp]
	\centering
	\caption{Mean and covariance contributions to task risk.}
	\label{tab:task_risk_component_summary}
	\begin{tblr}{
			width=\columnwidth,
			colspec={X[0.8] X[0.8] X[0.8] X[0.8]},
			row{1}={font=\bfseries},
			cells={font=\footnotesize}
		}
		\toprule
		Component                   & Adjoint/BF & Tikhonov & TSVD   \\
		\midrule
		Mean average fraction       & 0.5675     & 0.5691   & 0.5639 \\
		Covariance average fraction & 0.4325     & 0.4309   & 0.4361 \\
		Mean median fraction        & 0.8024     & 0.8058   & 0.7955 \\
		\bottomrule
	\end{tblr}
\end{table}

\begin{figure}[htbp]
	\centering
	\includegraphics[width=\columnwidth]{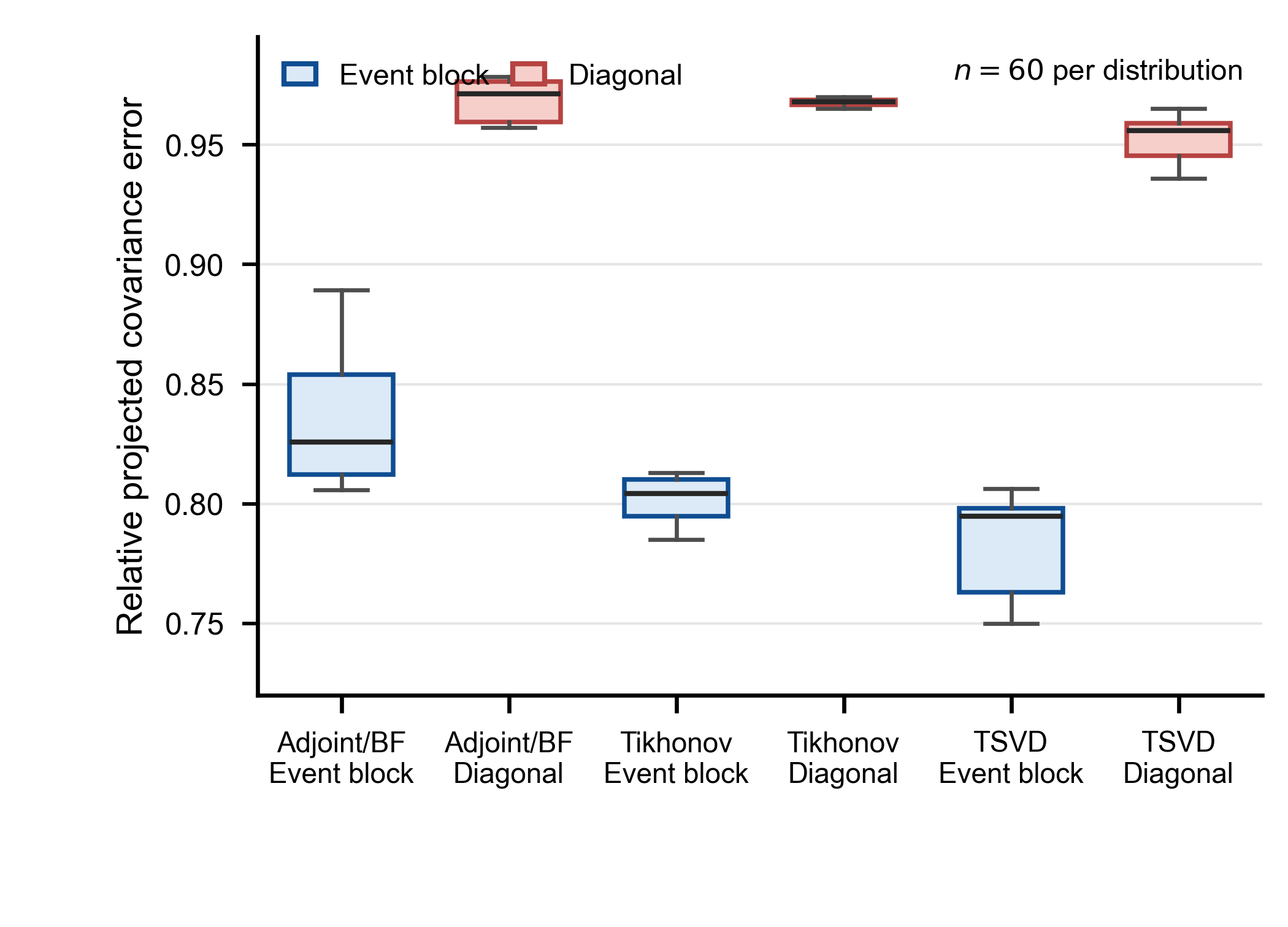}
	\caption{Task-domain consequences of event-block and diagonal covariance approximations.}
	\label{fig:task_covariance_approximation}
\end{figure}

Event-block projected covariance errors had medians $0.8257$, $0.8044$, and $0.7947$ for Adjoint/BF, Tikhonov, and TSVD. The corresponding diagonal errors were $0.9714$, $0.9679$, and $0.9560$. Absolute numerators, full projected covariance norms, and operator norms were also retained because relative errors with a small denominator cannot be used to rank operators. These projected covariance results provide the experimental check of \cref{prop:task_covariance_projection}, including the full-to-reduced covariance difference. The projected covariance results are shown in \cref{fig:task_covariance_approximation} and summarized in \cref{tab:task_covariance_scale}.

\begin{table}[htbp]
	\centering
	\caption{Median projected covariance errors for the two covariance approximations.}
	\label{tab:task_covariance_scale}
	\footnotesize
	\begin{tblr}{
			width=\columnwidth,
			colspec={X[1.2] X[1.0] X[1.0] X[1.0]},
			row{1}={font=\bfseries},
			cells={font=\footnotesize}
		}
		\toprule
		Approximation & Adjoint/BF & Tikhonov & TSVD   \\
		\midrule
		Event-block   & 0.8257     & 0.8044   & 0.7947 \\
		Diagonal      & 0.9714     & 0.9679   & 0.9560 \\
		\bottomrule
	\end{tblr}
\end{table}

\begin{figure*}[htbp]
	\centering
	\subfloat[Clean target]{\includegraphics[width=0.30\textwidth]{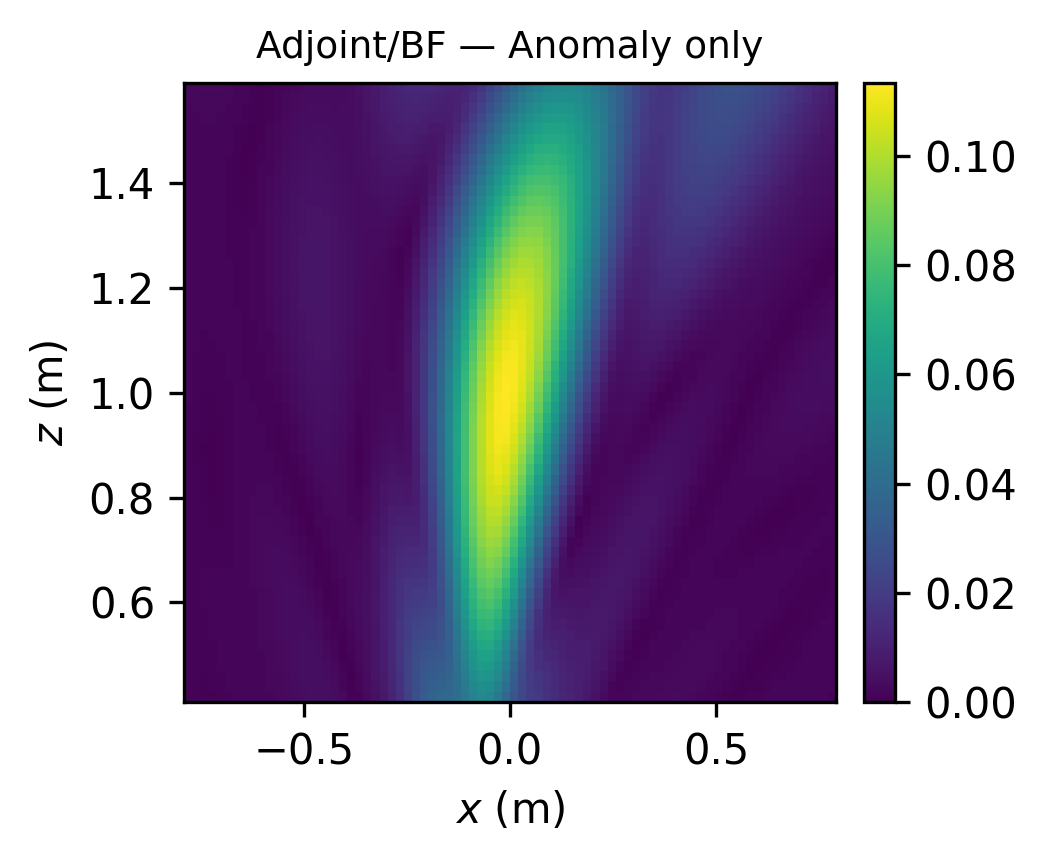}}
	\subfloat[Residual only]{\includegraphics[width=0.30\textwidth]{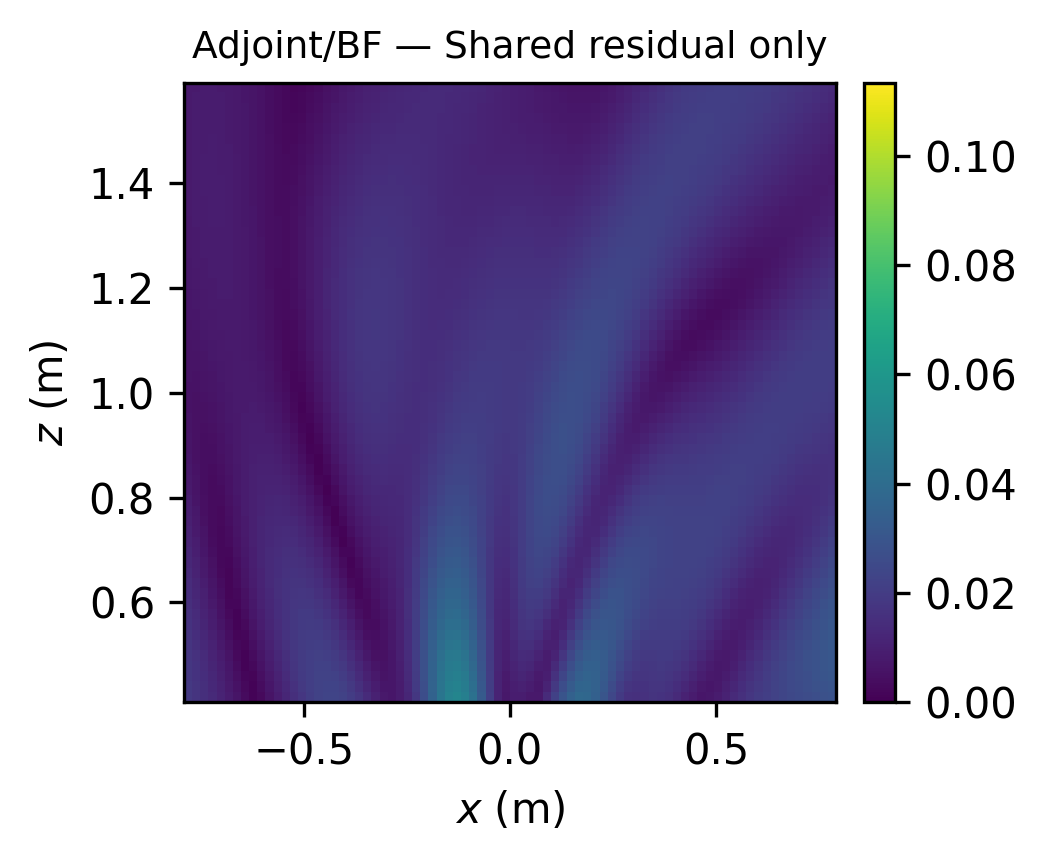}}
	\subfloat[Target plus residual]{\includegraphics[width=0.30\textwidth]{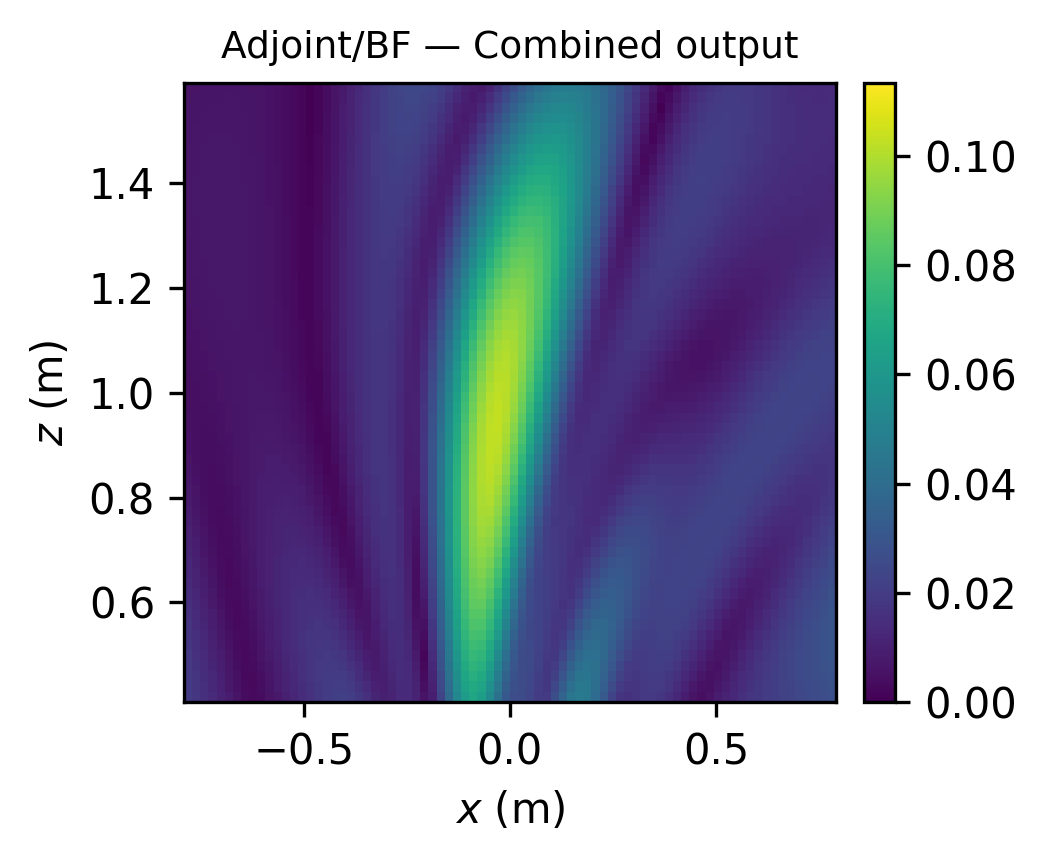}}\\
	\subfloat[Tikhonov clean target]{\includegraphics[width=0.30\textwidth]{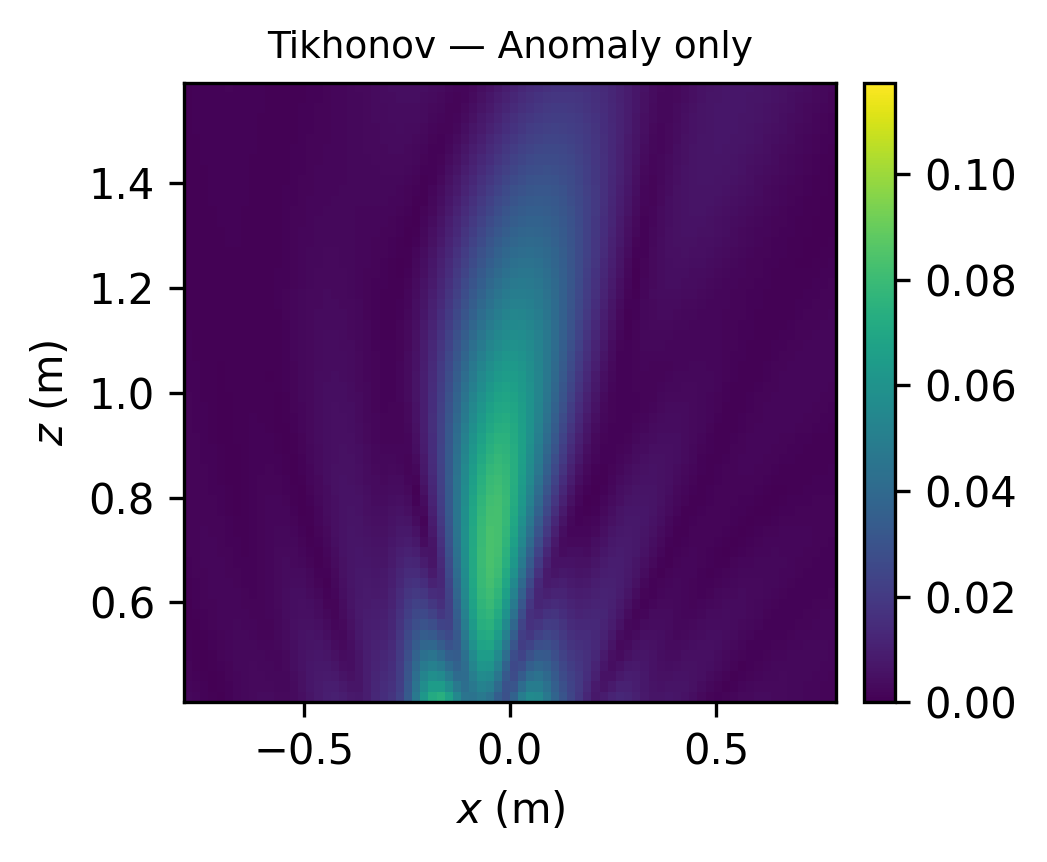}}
	\subfloat[Tikhonov residual only]{\includegraphics[width=0.30\textwidth]{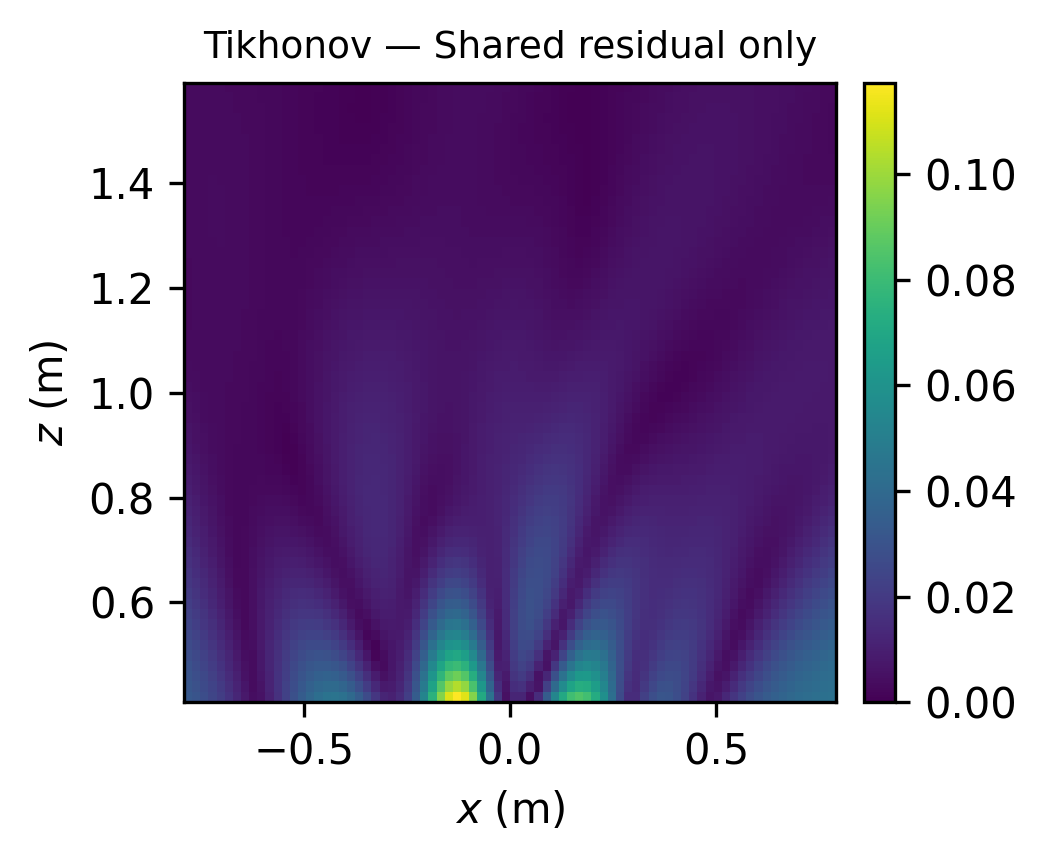}}
	\subfloat[Tikhonov target plus residual]{\includegraphics[width=0.30\textwidth]{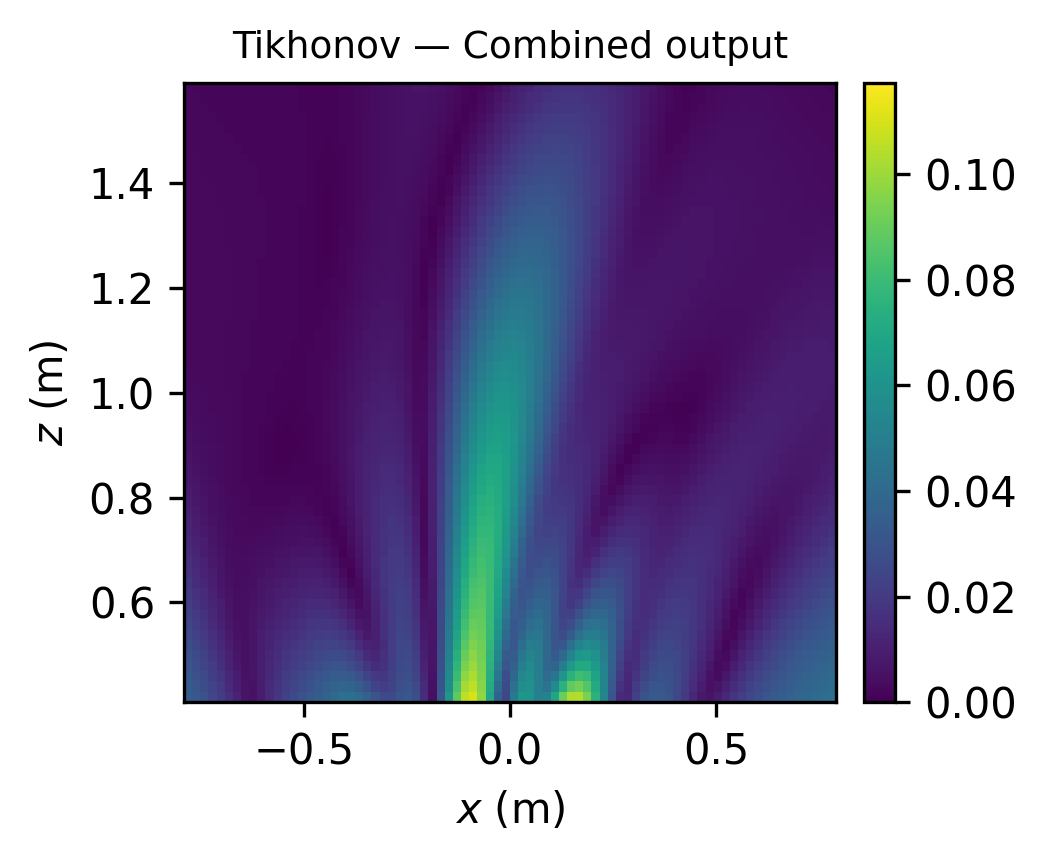}}\\
	\subfloat[TSVD clean target]{\includegraphics[width=0.30\textwidth]{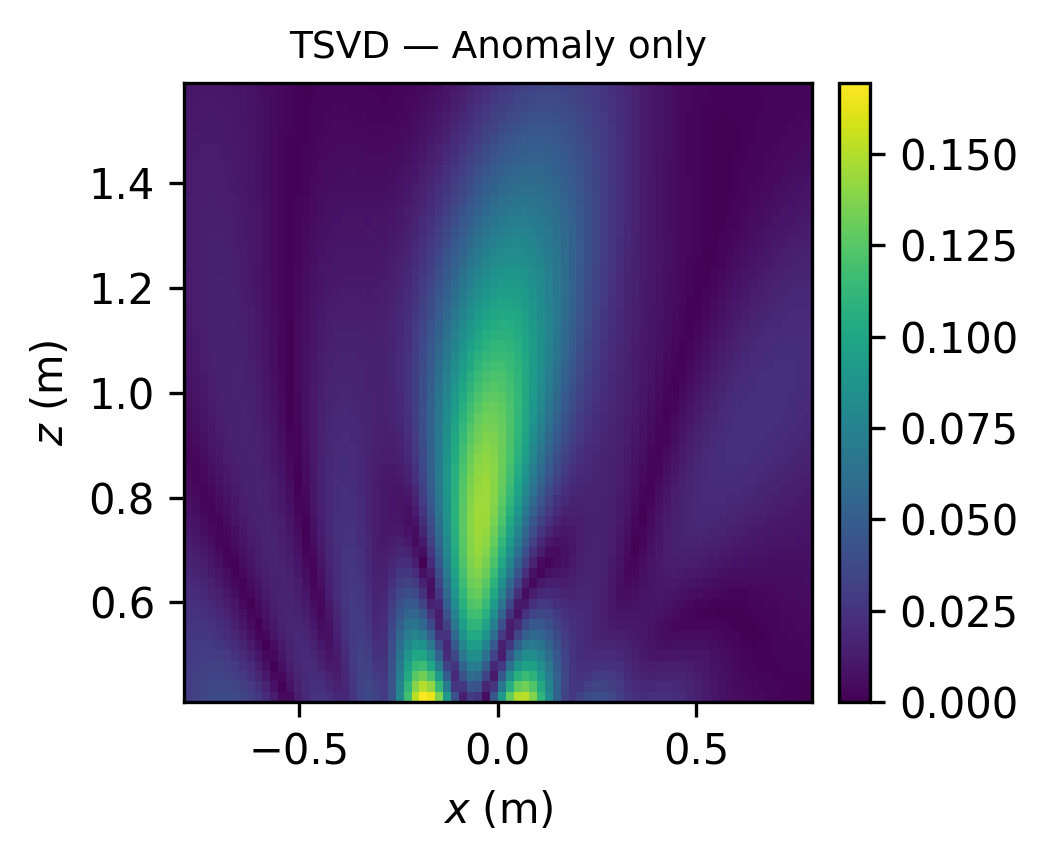}}
	\subfloat[TSVD residual only]{\includegraphics[width=0.30\textwidth]{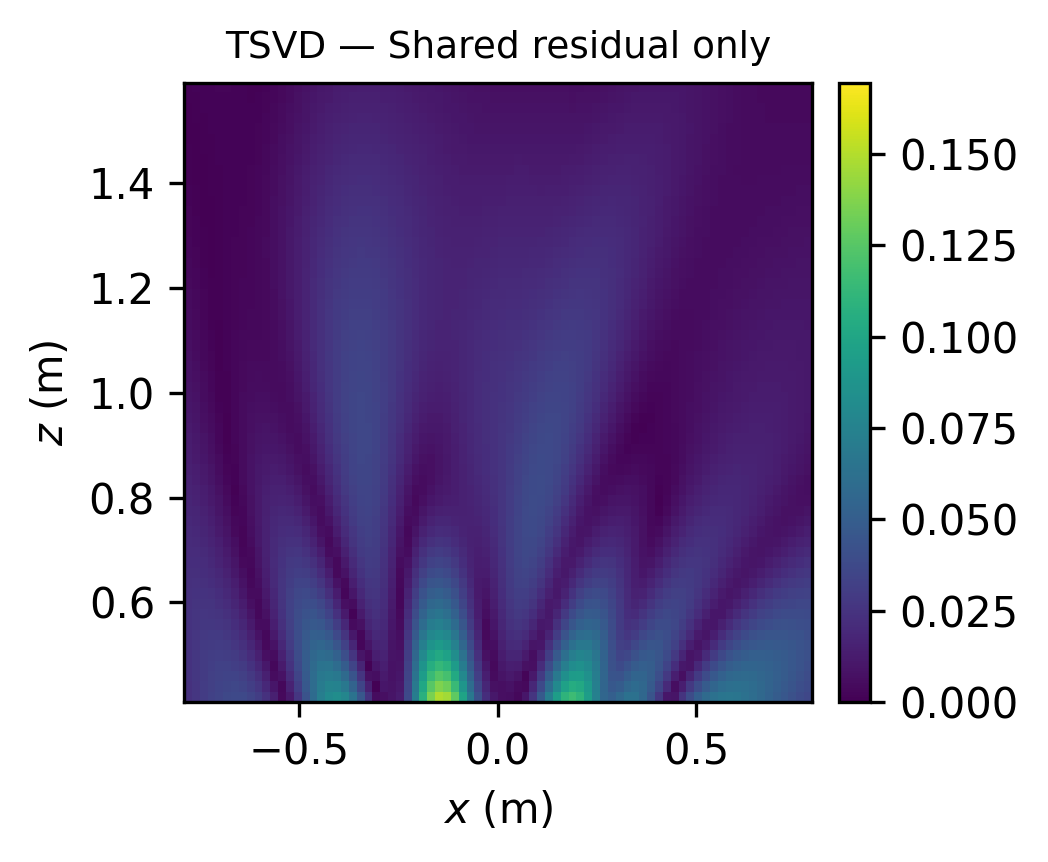}}
	\subfloat[TSVD target plus residual]{\includegraphics[width=0.30\textwidth]{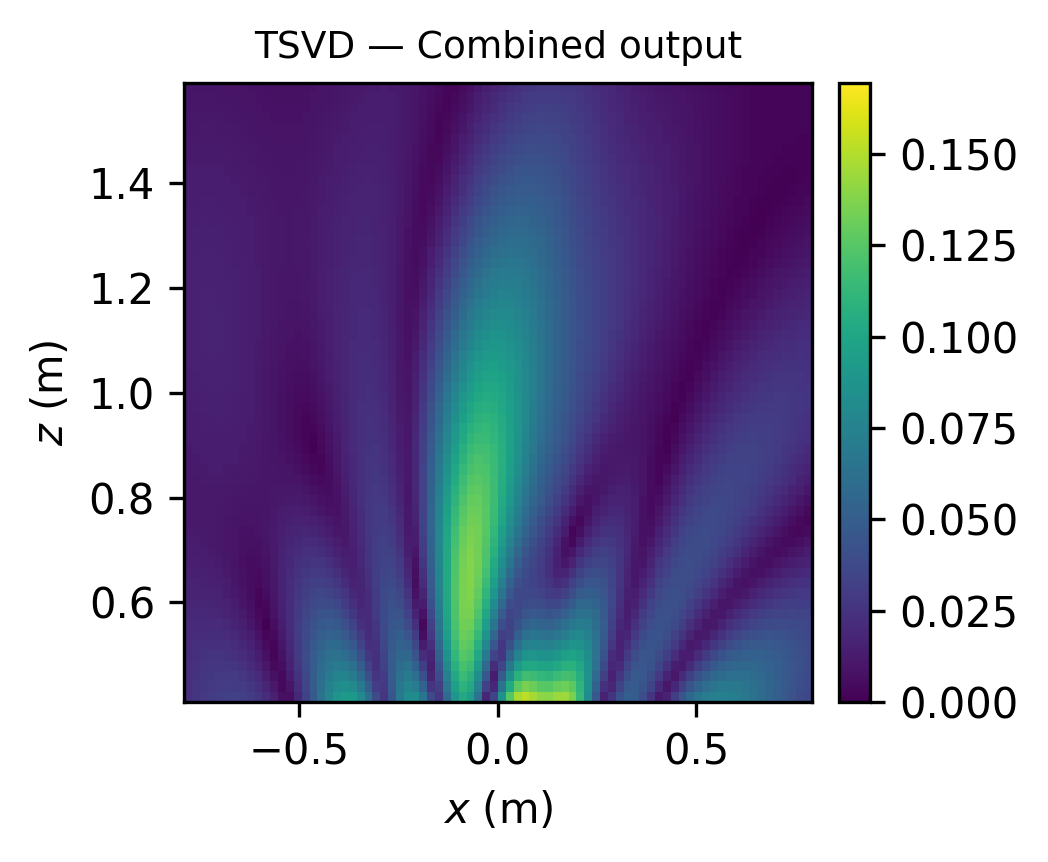}}
	\caption{Fixed-task spatial consequences of the shared residual.}
	\label{fig:analytical_task_spatial_consequences}
\end{figure*}

The spatial maps in \cref{fig:analytical_task_spatial_consequences} show clean-target, residual-only, and target-plus-residual outputs for the three ports. Each row was independently amplitude-normalized, so comparisons are valid within a row; amplitudes are not compared across rows. These are fixed-task spatial maps. They are not B-scans, measured images, target-detectability results, or evidence that all reconstruction algorithms produce the same artifact.

\begin{figure}[htbp]
	\centering
	\includegraphics[width=0.48\columnwidth]{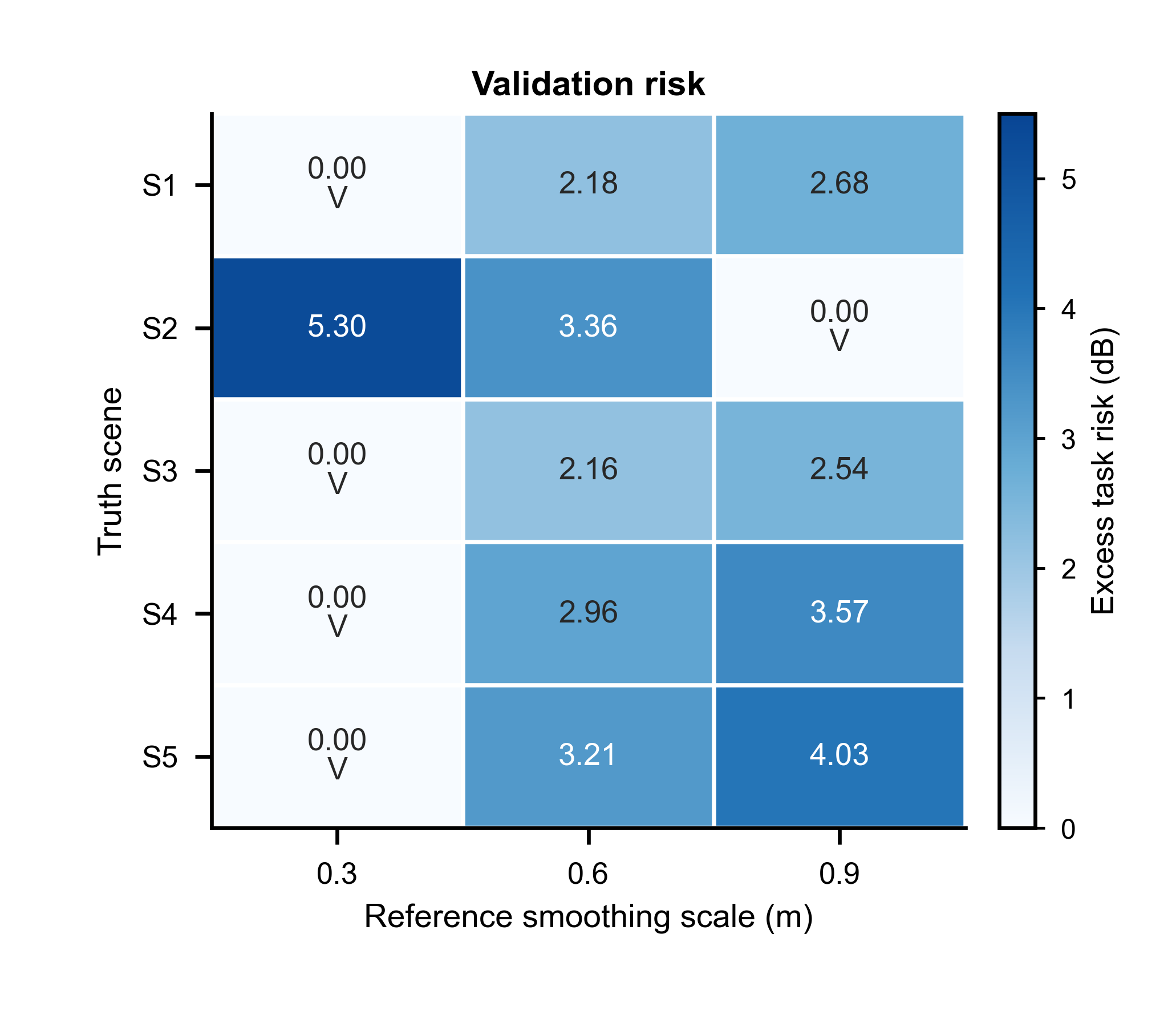}
	\includegraphics[width=0.48\columnwidth]{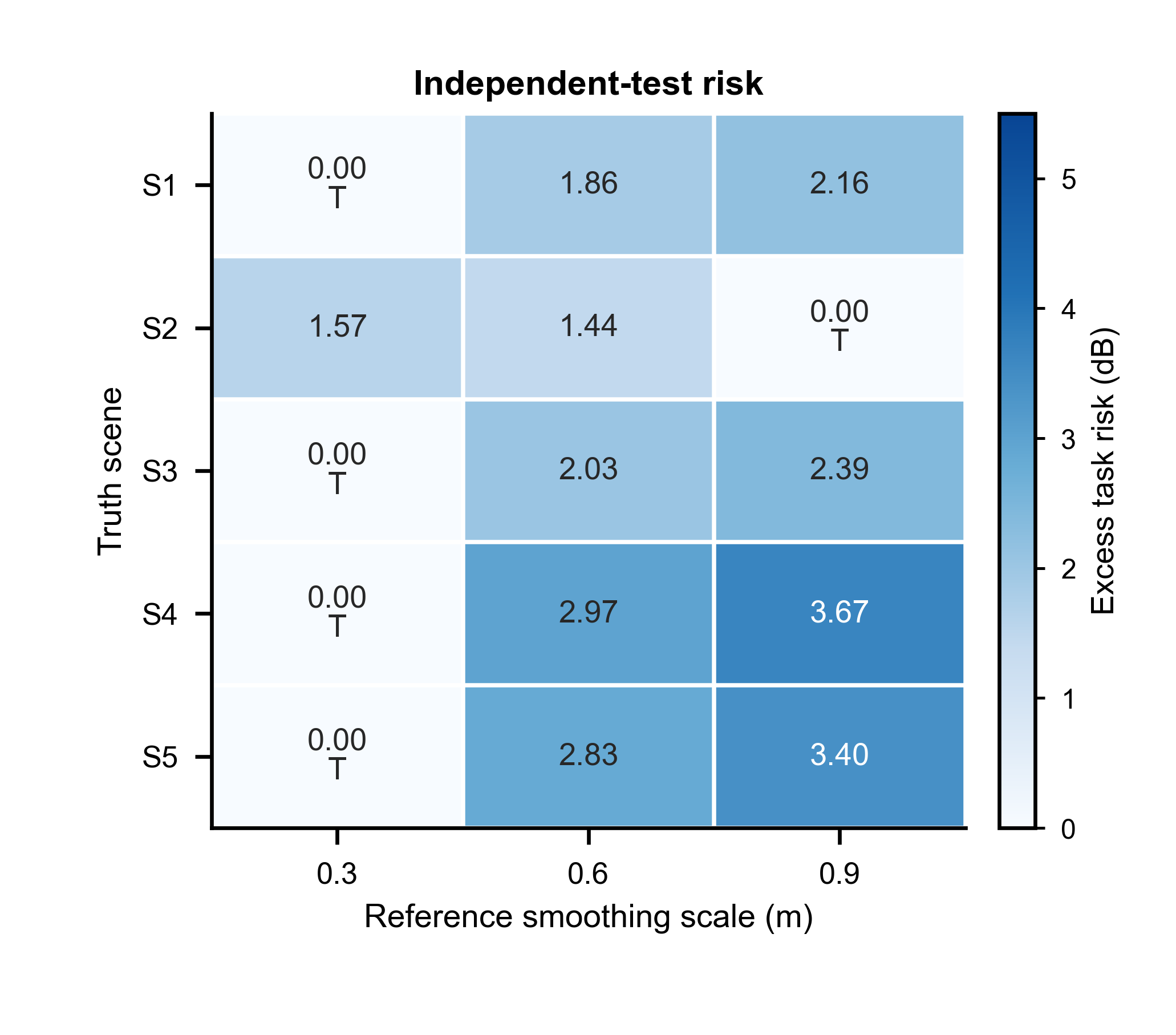}
	\caption{Validation and test behavior of the finite reference-scale screening interface.}
	\label{fig:reference_scale_screening}
\end{figure}

The finite screening protocol used physically admissible candidates and retained only candidates within the analytical first-order operational domain. Training samples 1--256 estimated moments, validation samples 257--384 selected a candidate, and test samples 385--512 evaluated it independently. Ties within $1\%$ were resolved by lower operator gain. Among three spatial correlation scales, validation and test selected $0.3~\si{m}$ for S1, S3, S4, and S5 and $0.9~\si{m}$ for S2. Relative to the $0.6~\si{m}$ baseline, test risk decreased by $28.1$--$49.6\%$. The scale-selection behavior is shown in \cref{fig:reference_scale_screening}.

For ten candidates and the three task ports, validation and test optima agreed in $12$ of $15$ scene--task combinations. The maximum normalized regret was $0.0086$. Adjoint/BF had zero regret in five of five cases, Tikhonov had regret $0.0017$ in four of five cases, and TSVD had regret $0.0086$ in three of five cases. The lowest-batch selection frequencies were $1$, $0.875$, and $0.625$. The screening summary is listed in \cref{tab:reference_selection_summary}.

\begin{table}[htbp]
	\centering
	\caption{Finite reference-screening summary.}
	\label{tab:reference_selection_summary}
	\footnotesize
	\begin{tblr}{
			width=\columnwidth,
			colspec={X[1.4] X[2.0]},
			row{1}={font=\bfseries},
			cells={font=\footnotesize}
		}
		\toprule
		Quantity                           & Result                           \\
		\midrule
		Validation--test optimum agreement & $12/15$ scene--task combinations \\
		Maximum normalized regret          & $0.0086$                         \\
		Adjoint/BF                         & $5/5$ zero-regret cases          \\
		Tikhonov                           & $4/5$ cases; regret $0.0017$     \\
		TSVD                               & $3/5$ cases; regret $0.0086$     \\
		\bottomrule
	\end{tblr}
\end{table}

\begin{figure}[htbp]
	\centering
	\subfloat[Selected reference effect]{\includegraphics[width=0.48\columnwidth]{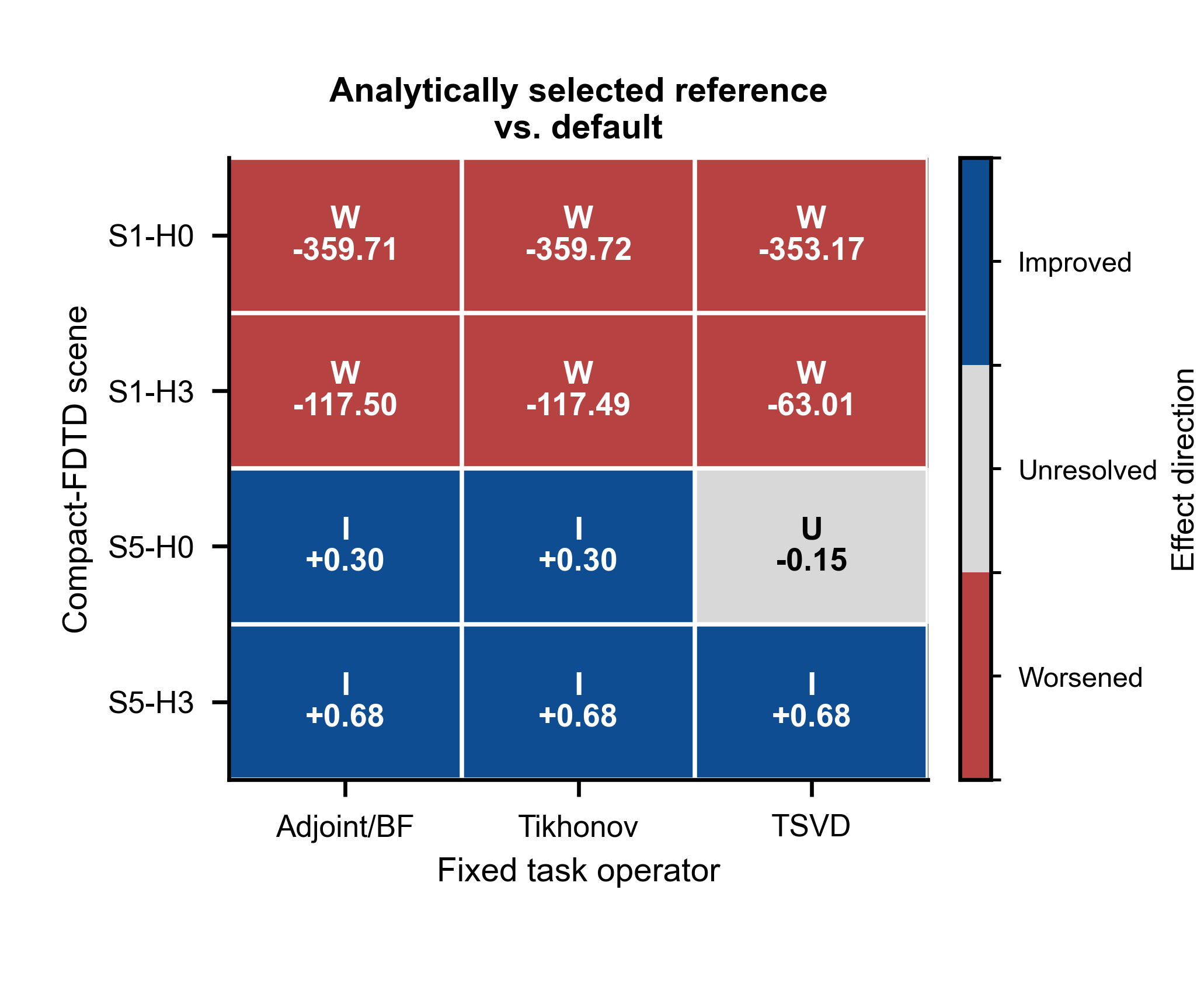}\label{fig:fdtd_selected_reference_effect_direction}}
	\hfill
	\subfloat[Wrong reference effect]{\includegraphics[width=0.48\columnwidth]{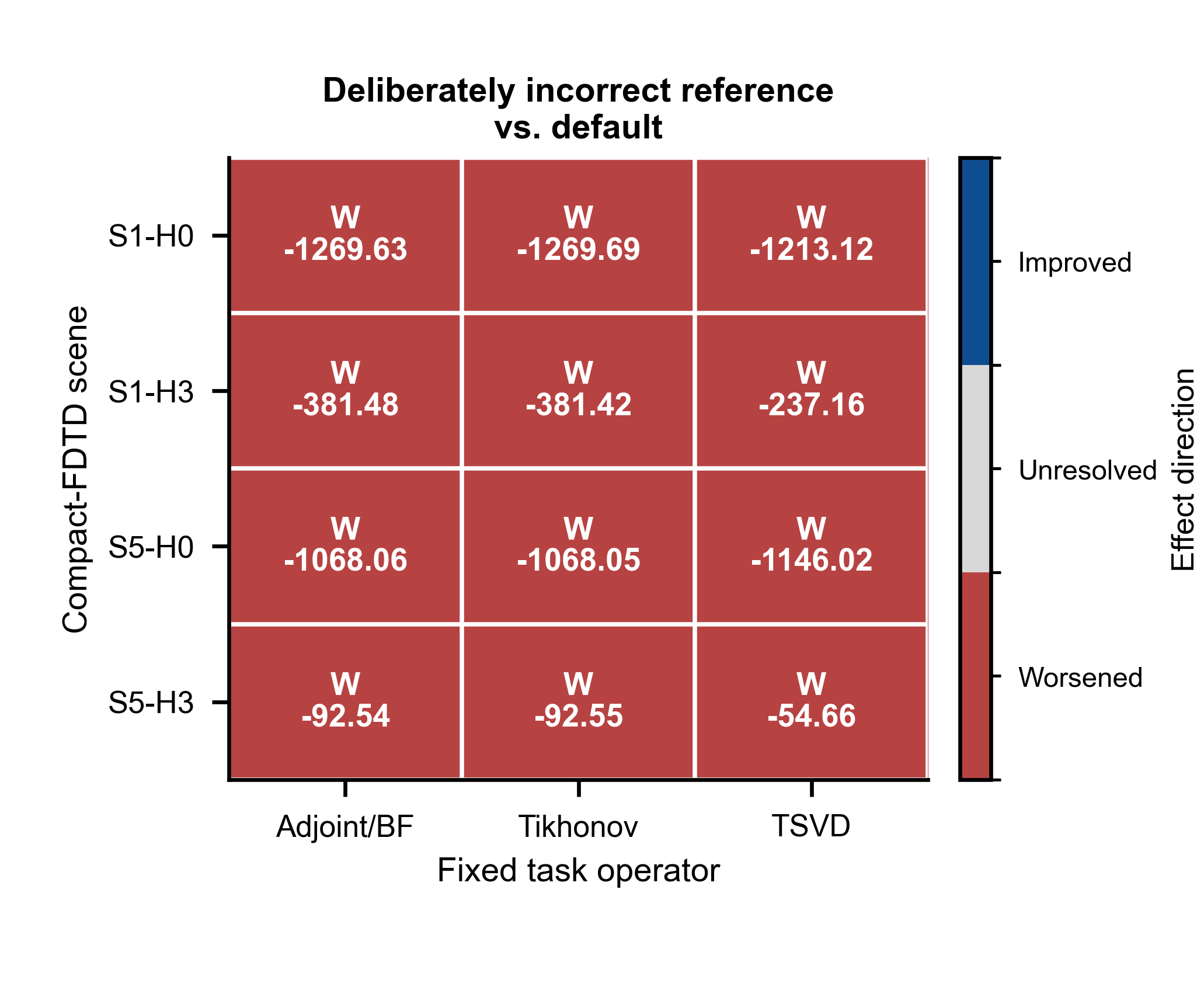}\label{fig:fdtd_wrong_reference_effect_direction}}
	\caption{Finite full-wave direction checks for selected and wrong reference states.}
	\label{fig:fdtd_reference_effect_direction}
\end{figure}

The finite full-wave direction check gave mixed outcomes. For the selected analytical reference, five conditions improved task energy, six worsened, and one was unresolved. The deliberately wrong reference worsened all twelve tested conditions. This result supports conditional risk screening under the tested candidate library and task ports. The selected and wrong-reference directions are shown in \cref{fig:fdtd_reference_effect_direction}.

% \FloatBarrier

\section{Conclusion}\label{sec:conclusion}

This study formulated single-snapshot FDA-MIMO-GPR propagation in complex media through the standard DBA route for incremental scattering. The selected reference medium provided the working state for separating the anomaly-free physical background from a local anomaly. The reference--background parameter residual then entered the fixed first-order propagation model as a shared residual source.

The theory distinguished the first-order shared response from the exact full-wave background difference and its modeling remainder. It derived observation-domain moments, task-domain moment and covariance-projection identities, and exact-to-first-order discrepancy transfer for any fixed linear task map. The experiments instantiated these relations using Adjoint/BF, Tikhonov, and TSVD. The formal statements remain tied to the fixed reference, stated covariance models, and linear task formulation.

The experimental validation used analytical ensembles and finite full-wave simulations as complementary layers. The analytical results showed nonzero off-event covariance, spectral concentration, task-dependent residual energy, and structured residual-only spatial maps. Frequency states and transmit--frequency assignments reorganized these features through the propagation operator. The common reference--background residual remained their physical source. The mismatch sweep showed increasing first-order discrepancy, while finite full-wave checks showed that the remainder can approach or exceed the complete task response under some tested conditions.

The finite reference-screening interface ranked candidates for a fixed task, normalization, candidate library, and first-order validity domain. Validation and test optima agreed in 12 of 15 scene--task combinations, with a maximum normalized regret of 0.0086. The finite full-wave direction checks produced five improvements, six degradations, and one unresolved case for the analytically selected references, whereas the deliberately wrong reference degraded all twelve tested conditions. The interface therefore provides conditional candidate ranking rather than guaranteed task improvement, material identification, or parameter recovery.

The theoretical formulation uses a fixed reference kernel and a first-order propagation model. Its quantitative interpretation is restricted by the constitutive, propagation, interface, and scattering effects collected in the remainder. The validation uses controlled Cole--Cole ensembles, finite full-wave simulations, idealized sources and polarization, a restricted assignment family, and the reported task ports. Measured GPR data, general three-dimensional heterogeneous-strata validation, realistic antenna coupling and calibration, and a nonlinear strong-scattering stability analysis are outside the current study.

Future work should extend the model to three-dimensional heterogeneous media, realistic antenna coupling and calibration uncertainty, measured FDA-MIMO-GPR acquisitions, broader frequency and coding designs, and adaptive, Bayesian, or detection-oriented task operators. These extensions can test which parts of the residual covariance and task-transfer structure persist beyond the present fixed-reference validation domain.

\bibliographystyle{IEEEtran}
\bibliography{ref}

\end{document}